\documentclass[aps]{revtex4}
\usepackage{eurosym}
\usepackage{amsfonts}
\usepackage{amsmath}
\usepackage{amssymb,epsf}
\usepackage{color}
\usepackage{graphicx}
\usepackage{natbib}
\usepackage{float}
\usepackage{caption}
\usepackage{subfig}
\usepackage{epstopdf}

\begin{document}

\title{Three-dimensional AdS black holes in massive-power-Maxwell theory}
\author{B. Eslam Panah$^{1,2,3}$\footnote{
email address: eslampanah@umz.ac.ir}, K. Jafarzade$^{1,2}$\footnote{
email address: khadije.jafarzade@gmail.com}, A. Rinc\'{o}n$^{4,5}$\footnote{
email address: angel.rincon@ua.es}}
\affiliation{$^{1}$ Department of Theoretical Physics, Faculty of Basic Sciences,
University of Mazandaran, P. O. Box 47416-95447, Babolsar, Iran\\
$^{2}$ ICRANet-Mazandaran, University of Mazandaran, P. O. Box 47416-95447,
Babolsar, Iran\\
$^{3}$ICRANet, Piazza della Repubblica 10, I-65122 Pescara, Italy\\
$^{4}$Departamento de F\'{\i}sica Aplicada, Universidad de Alicante, Campus
de San Vicente del Raspeig, E-03690 Alicante, Spain\\
$^{5}$Sede Esmeralda, Universidad de Tarapac\'{a}, Avda. Luis Emilio
Recabarren 2477, Iquique, Chile}

\begin{abstract}
Recently, it was shown that the power-Maxwell (PM) theory could remove the
singularity of the electric field \cite{PM2}. Motivated by a great interest
in three-dimensional black holes and a surge of success in studying massive
gravity from both the cosmological and astrophysical points of view, we
investigate three-dimensional black hole solutions in de Rham, Gabadadze,
and Tolley (dRGT) massive theory of gravity in the presence of PM
electrodynamics. First, we extract exact three-dimensional solutions in this
theory of gravity. Then we study the geometrical properties of these
solutions. Calculating conserved and thermodynamic quantities, we check the
validity of the first law of thermodynamics for these black holes. We also
examine the stability of these black holes in the context of the canonical
ensemble. We continue calculating this kind of black hole's optical
features, such as the photon orbit radius, the energy emission rate,
and the deflection angle. Considering these optical quantities, finally, we
analyze the effective role of the parameters of models on them.
\end{abstract}

\maketitle

\section{Introduction}

Considering General Relativity (GR) in three-dimensional spacetime, Banados,
Teitelboim, and Zanelli (BTZ) have found a black hole solution \cite{BTZ},
which is known as a BTZ black hole. The study of these black holes opened
different aspects of physics in three-dimensional spacetime, such as the
existence of specific relations between the BTZ black holes and effective
action in string theory \cite{Witten1998,Lee1999,Larranaga2008}, providing
simplified machinery for studying different features of black holes such as
thermodynamic ones \cite{ThBTZI,ThBTZII,ThBTZIII,ThBTZIV,ThBTZVI,ThBTZIX},
contributing to our understanding of gravitational systems and their
interactions in lower dimensions \cite{Witten2007}, the possible existence
of gravitational Aharonov-Bohm effect due to the non-commutative BTZ black
holes \cite{Anacleto2015}, AdS/CFT correspondence \cite%
{Emparan2000,Carlip2005,Setare2007}, quantum aspect of three-dimensional
gravity, entanglement, and quantum entropy \cite%
{quBTZI,quBTZII,quBTZIII,quBTZIV,quBTZVI}, holographic aspects of BTZ black
hole solutions \cite{HoloBTZI,HoloBTZII,HoloBTZIII}, and anti-Hawking
phenomena of BTZ black holes \cite{antiBTZI,antiBTZII}. It is essential to
point out that gravity in three-dimensional spacetime is a vibrant field of
research in part because the absence of propagating degrees of freedom makes
things more straightforward than in four dimensions, in particular, when
dealing with the challenge of formulating a quantization of this theory.
Also, the BTZ black hole is currently a seminal toy model to study different
effects beyond GR. On the other hand, by considering various theories of
gravity coupled with different matter fields, three-dimensional black hole
solutions and their thermodynamic properties have been studied in literature 
\cite{NBTZ1,NBTZ5,NBTZ6,NBTZ7,NBTZ10,NBTZ12,NBTZ13,NBTZ14,NBTZ16,NBTZ18}.

One of the most challenging problems of modern cosmology is related to the
fact that our Universe is expanding with acceleration. Some candidates have
been proposed to explain this acceleration, such as the existence of a
positive cosmological constant, dark energy, and modified theories of
gravity. Among different candidates of modified theories of gravity, massive
theories of gravity have attracted much attention lately due to a wide
variety of motivations in various aspects of physics \cite{Hinterbichler,de
RhamI}. From a cosmological point of view, one can point out interesting
features such as describing the accelerating expansion of our Universe
without requiring any dark energy \cite{AkramiI,AkramiII}, explaining the
current observations related to dark matter \cite{BabichevI,BabichevII},
suitable description of rotation curves of the Milky Way, spiral galaxies,
and low surface brightness galaxies \cite{Panpanich}. The most important
achievements in the astrophysics context are: the existence of white dwarfs
more than the Chandrasekhar limit \cite{White} and massive neutron stars
with a maximum mass more than three times the solar mass \cite{Neutron}. To
name a few in the point of black hole physics, one can mention: the
existence of van der Waals-like behavior in extended phase space for
non-spherical black holes \cite{MassBHI,MassBHII}, triple points and N-fold
reentrant phase transitions \cite{Dehghani}, the existence of a remnant for
a black hole which may help to ameliorate the information paradox \cite%
{RemnantI,RemnantII}, etc.

In recent years, a new version of the theory of massive gravity has been
proposed by de Rham, Gabadadze, and Tolley (dRGT) \cite{dRGTII,dRGTIII},
which is known as dRGT massive gravity. The dRGT massive gravity's action
contains a nonlinear interaction term that admits the Vainshtein mechanism,
and it is free from van Dam-Veltman-Zakharov discontinuity \cite%
{vDVZI,vDVZII}, and Boulware-Deser ghost \cite{BDI,BDII} in arbitrary
dimensions (which appears in Fierz-Pauli theory of massive gravity \cite%
{Fierz}). Notably, the dRGT theory of massive gravity requires a fiducial
reference metric ($f_{\mu \nu }$) in addition to the dynamical metric ($%
g_{\mu \nu }$) to define a mass term for graviton by introducing some
non-derivative potential terms ($\mathcal{U}_{i}$). Also, modification in
the introduced reference metric leads to a special family of dRGT massive
gravity \cite{de RhamI}. So, different reference metrics could lead to a
variety of new solutions. In this regard, it was shown that the dRGT massive
gravity is ghost-free by considering different reference metrics such as
Minkowski and degenerate (singular) reference metrics (see Refs. \cite%
{GhmassII,GhmassIII,GhmassIV}, for more details). Asymptotically flat and
(A)dS black holes in the context of massive gravity have been obtained by
considering the flat (Minkowski) reference metric or on a degenerate
(spatial) and singular reference metric \cite%
{BHmassI,BHmassII,BHmassIII,BHmassIV,BHmassV,BHmassVI}.

One of the interesting cases of theories of massive gravity is related to
the AdS black hole solutions with the degenerate (spatial) reference metric,
which is singular and has important applications in gauge/gravity duality
(see Ref. \cite{BHmassVI}, for more details). In this theory of massive
gravity, graviton may behave like a lattice and exhibit a Drude peak \cite%
{BHmassVI}. It was indicated that this theory of massive gravity is stable
and ghost-free\ \cite{GhmassIII}. Black hole solutions in the context of
this massive gravity have been studied in Refs. \cite{VeghBHI,VeghBHII}.
Study on black holes in this theory has attracted extensive attention
recently, ranging from heat engine and Joule-Thomson expansion \cite%
{Heat,Heat2}, quasinormal modes \cite{quasinormalI,quasinormalII}, van der
Waals-like phase transition \cite{MassBHI,MassBHII,Dehghani,Heat,VdW},
reentrant phase transitions and triple points \cite{reeentraint}, phase
transition and entropic force \cite{entropyF}, thermodynamics and
geometrical thermodynamics \cite{GTI,GTII,GTIII,GTIV,GTV}, and also the
existence correspondence between black hole solutions of conformal and
massive theories of gravity \cite{ConformalMassive}. In the present work,
motivated by the interesting properties of three-dimensional black hole
solutions, we study BTZ black holes in this specific massive gravity theory
which has a lot of applications in AdS/CMT \cite{CaiMassive,Hendi2017ab},
QCD \cite{Neutron}, quantum information \cite{Ghodrati2019} and studies of
black hole information paradox.

Another fascinating subject that has gained significant attention is the
coupling of theories of gravity with nonlinear electrodynamics (NED). The
power-Maxwell (PM) theory is one of the interesting and special classes of
NED that Hassaine and Martinez presented in 2007 \cite{PM1}. Recently, it
was shown that PM theory, similar to Born-Infeld theory, could remove the
singularity of the electric field at the origin \cite{PM2}. The Lagrangian
of PM theory is an arbitrary power of Maxwell Lagrangian, where it is
invariant under the conformal transformation $g_{\mu \nu }\rightarrow
\Omega^{2}g_{\mu \nu }$ (where $g_{\mu \nu }$ is metric tensor) and $%
A_{\mu}\rightarrow A_{\mu }$, see Ref. \cite{PM4}, for more details. It is
worth mentioning that the PM theory reduces to linear Maxwell theory when
the power of Maxwell is unit \cite{PM4}. Another marvelous feature of PM is
related to its conformal invariance. As we know, the Maxwell action enjoys
conformal invariance in four dimensions, but it does not possess this
symmetry in higher dimensions. However, the Lagrangian of PM theory extends
the conformal invariance in higher dimensions if the power is chosen as $%
s=\left( \text{dimensions of spacetime}\right) /4$ (where $s$ is the power
of PM theory). This leads to black hole solutions, which are inverse square
electric fields in arbitrary dimensions (the so-called Coulomb law). In this
regard, some interesting properties of black hole solutions coupled with the
PM theory have been studied in Refs. \cite%
{BHPN1,BHPN2,BHPN3,BHPN4,BHPN5,BHPN6,BHPN7}. Generalization of GR with a
massive spin$-2$ field Lagrangian minimally coupled to a PM $U(1)$ gauge
field in four and higher dimensional spacetime have been investigated in
Refs. \cite{Dehghani,AcPMI}, which led to some novel and interesting
properties of black hole physics.

Taking into account the mentioned motivations, in this paper, we will
extract three-dimensional black hole solutions by considering three
generalizations, a massive spin$-2$ field, a PM $U(1)$ gauge field, and the
cosmological constant to the Einstein-Hilbert Lagrangian. Then we study
their properties from various perspectives.

This paper is organized as follows: In Sec. II, we consider the
three-dimensional action of Einstein-dRGT massive gravity in the presence of
PM electrodynamics and obtain field equations. Solving the field equations
analytically, we obtain exact three-dimensional solutions in the PM-dRGT
massive gravity and discuss the main properties of the solution in Sec. III.
Thermodynamic behaviors and the stability of the black hole solutions are
investigated in Sec. IV. Moreover, it is shown that the first law is valid
for the obtained solutions. In Sec. V, we determine the null geodesics equations and radius of the photon orbit. We find the
allowed regions of the model parameters to have acceptable optical behavior.
The energy emission rate and deflection angle for this type of solution are
investigated in this section. Finally, We end the paper with remarkable
results in Sec. VI.

\section{Basic Equations}

\label{SecII} The three-dimensional action of Einstein-dRGT massive gravity
coupled with the PM nonlinear electrodynamics is given by \cite{AcPMI} 
\begin{equation}
\mathcal{I}=-\frac{1}{16\pi }\int d^{3}x\sqrt{-g}\left[ \mathcal{R}-2\Lambda
+\left( -\mathcal{F}\right) ^{s}+m_{g}^{2}\varepsilon _{i}\mathcal{U}%
_{i}(g,f)\right] ,  \label{Action}
\end{equation}%
where $\mathcal{R}$, and $\Lambda $\ are the scalar curvature and the
cosmological constant, respectively. Also, $\mathcal{F}=F_{\mu \nu }F^{\mu
\nu }$\ is the Maxwell invariant (where $F_{\mu \nu }=\partial _{\mu }A_{\nu
}-\partial _{\nu }A_{\mu }$\ and $A_{\mu }$\ are the Faraday tensor and the
gauge potential, respectively). Here, $s$ is related to the power of PM
theory. It is straightforward to show that the term $\left( -\mathcal{F}%
\right) ^{s}$ reduces to the standard Maxwell Lagrangian for $s=1$. In the
above action, $m_{g}$ and $f$ are related to the graviton mass and a fixed
symmetric tensor, respectively. Note that $\mathbf{\varepsilon }_{i}$'s are some constants and also $\mathcal{U}_{i}$'s are self-interaction
potentials of graviton constructed from the building blocks $\mathcal{K}%
_{\nu }^{\mu }=\sqrt{g^{\mu \alpha }f_{\alpha \nu }}$ which can be written
as follows 
\begin{eqnarray}
\mathcal{U}_{1} &=&\left[ \mathcal{K}\right] ,\;\;\;\;\;\mathcal{U}_{2}=%
\left[ \mathcal{K}\right] ^{2}-\left[ \mathcal{K}^{2}\right] ,\;\;\;\;\;%
\mathcal{U}_{3}=\left[ \mathcal{K}\right] ^{3}-3\left[ \mathcal{K}\right] %
\left[ \mathcal{K}^{2}\right] +2\left[ \mathcal{K}^{3}\right] ,  \notag \\
&&\mathcal{U}_{4}=\left[ \mathcal{K}\right] ^{4}-6\left[ \mathcal{K}^{2}%
\right] \left[ \mathcal{K}\right] ^{2}+8\left[ \mathcal{K}^{3}\right] \left[ 
\mathcal{K}\right] +3\left[ \mathcal{K}^{2}\right] ^{2}-6\left[ \mathcal{K}%
^{4}\right] .  \notag
\end{eqnarray}%
The square root in $\mathcal{K}$ means $\left( \mathcal{K}\right) _{\lambda
}^{\mu }\left( \mathcal{K}\right) _{\nu }^{\lambda }=\mathcal{K}_{\nu }^{\mu
}$ and the rectangular brackets denote traces, $[\mathcal{K}]=\mathcal{K}%
_{\mu }^{\mu }$. Taking into account the action (\ref{Action}) and using
the variational principle, we can extract the field equations related to the
gravitation and gauge fields as \cite{AcPMI}%
\begin{eqnarray}
G_{\mu \nu }+\Lambda g_{\mu \nu }+m_{g}^{2}\chi _{\mu \nu } &=&\frac{1}{2}%
g_{\mu \nu }\left( -\mathcal{F}\right) ^{s}+2s\left( -\mathcal{F}\right)
^{s-1}F_{\mu \rho }F_{\nu }^{\rho },  \label{Field equation} \\
&&  \notag \\
\partial _{\mu }\left( \sqrt{-g}\left( -\mathcal{F}\right) ^{s-1}F^{\mu \nu
}\right) &=&0,  \label{Maxwell equation}
\end{eqnarray}%
where $G_{\mu \nu }=R_{\mu \nu }-\frac{1}{2}g_{\mu \nu }R$, is Einstein's
tensor. Also $\chi _{\mu \nu }$ is the massive term with the
following form 
\begin{eqnarray}
\chi _{\mu \nu } &=&-\frac{\mathbf{\mathbf{\varepsilon }}_{1}}{2}\left( 
\mathcal{U}_{1}g_{\mu \nu }-\mathcal{K}_{\mu \nu }\right) -\frac{\mathbf{%
\mathbf{\varepsilon }}_{2}}{2}\left( \mathcal{U}_{2}g_{\mu \nu }-2\mathcal{U}%
_{1}\mathcal{K}_{\mu \nu }+2\mathcal{K}_{\mu \nu }^{2}\right) -\frac{\mathbf{%
\mathbf{\varepsilon }}_{3}}{2}(\mathcal{U}_{3}g_{\mu \nu }-3\mathcal{U}_{2}%
\mathcal{K}_{\mu \nu }+  \notag \\
&&6\mathcal{U}_{1}\mathcal{K}_{\mu \nu }^{2}-6\mathcal{K}_{\mu \nu }^{3})-%
\frac{\mathbf{\mathbf{\varepsilon }}_{4}}{2}(\mathcal{U}_{4}g_{\mu \nu }-4%
\mathcal{U}_{3}\mathcal{K}_{\mu \nu }+12\mathcal{U}_{2}\mathcal{K}_{\mu \nu
}^{2}-24\mathcal{U}_{1}\mathcal{K}_{\mu \nu }^{3}+24\mathcal{K}_{\mu \nu
}^{4}).  \label{massiveTerm}
\end{eqnarray}

\section{Black hole solutions}

\label{SecIII} In this section, we are interested in studying the
three-dimensional static black holes with (A)dS asymptotes in the presence
of PM theory and Einstein-dRGT-massive gravity. In this regard, we consider
the metric of three-dimensional static spacetime with the following explicit
form 
\begin{equation}
ds^{2}=-g(r)dt^{2}+g^{-1}(r)dr^{2}+r^{2}d\varphi ^{2},  \label{metric}
\end{equation}%
where $g(r)$ is an arbitrary function of the radial coordinate.

To obtain exact solutions, we should choose the reference metric. We
consider the following ansatz metric \cite{CaiMassive} 
\begin{equation}
f_{\mu \nu }=diag(0,0,c^{2}),  \label{f11}
\end{equation}%
where $c$ is a positive constant. Such a choice of reference metric
depends only on the spatial components; meaning that the diffeomorphism
invariance is preserved in the $t$ and $r$ coordinates (the breakdown of
diffeomorphism invariance leads to BD ghost. In the ADM language, the BD
ghost is a consequence of the absence of the Hamiltonian constraint \cite%
{Schmidt2012}. To solve this problem, one has to show that the Hamiltonian
constraint is preserved and thus eliminates the sixth degree of freedom (BD
ghost) \cite{Hassan041101}. In Ref. \cite{BHmassVI}, Vegh has presented a
version of this proof for the case of a degenerate reference metric $f_{\mu
\nu }=diag(0,0,1,1)$ and found that BD ghost eliminates because the
corresponding diffeomorphism remains unbroken in the $t$ and $r$
coordinates) but is broken in the spatial dimensions. One can imagine a more
general reference metric such that the diffeomorphism invariance in the $%
r-direction$ is broken. For instance, in four dimensions, to preserve
rotational invariance on the sphere and general time parametrization
invariance, a reference metric as $f_{\mu \nu
}=diag(0,1,c^{2},c^{2}sin^{2}\theta )$ was considered as a natural
ansatz \cite{Adams2015}. Another choice was a different generalization of $%
f_{\mu \nu }$, with $sin^{2}\theta f_{\theta \theta }=f_{\varphi \varphi
}=F(r)$ such that other components were zero \cite{Adams2015}. This can lead
to an ability to add arbitrary polynomial terms in $r$ to the emblackening
factor.

Using the metric ansatz (\ref{f11}), $\mathcal{U}(g,f)$ is given by \cite%
{CaiMassive} 
\begin{equation}
\mathcal{U}(g,f)=\frac{c}{r}.  \label{U}
\end{equation}

It is notable that in three-dimensional spacetime, the term $
m_{g}^{2}\varepsilon _{i}U_{i}(g,f)$ in the action (\ref{Action})
turns to $m_{g}^{2}\varepsilon U(g,f)$ or $\frac{m_{g}^{2}c\varepsilon }{r}$. Also, the massive term in Eq. (\ref%
{massiveTerm}) reduces to 
\begin{equation}
\chi _{\mu \nu }=\frac{-\varepsilon }{2}\left( g_{\mu \nu }\mathcal{U}(g,f)-%
\mathcal{K}_{\mu \nu }\right) .  \label{massiveTerm1}
\end{equation}

Since we are going to study electrically charged black holes, we consider a
radial electric field its related gauge potential is 
\begin{equation}
A_{\mu }=h\left( r\right) \delta _{\mu }^{t}.  \label{gauge
potential}
\end{equation}

Using the metric (\ref{metric}) and the PM field equation (\ref{Maxwell
equation}), one finds the following differential equation 
\begin{equation}
rh^{\prime \prime }(r)+\Psi _{1}=0,
\end{equation}%
where 
\begin{equation}
\Psi _{1}=\left\{ 
\begin{array}{ccc}
h^{\prime }(r) &  & s=1 \\ 
&  &  \\ 
2h^{\prime }(r) &  & s=\frac{3}{4} \\ 
&  &  \\ 
-h^{\prime }(r)-2srh^{\prime \prime }(r) &  & \text{otherwise}%
\end{array}%
\right. ,  \label{heq}
\end{equation}%
where the prime and double primes are the first and the second derivatives
versus $r$, respectively. It is easy to find the solution of Eq. (\ref{heq})
as 
\begin{equation}
h(r)=\left\{ 
\begin{array}{ccc}
\frac{q}{l}\ln \left( \frac{r}{l}\right) &  & s=1 \\ 
&  &  \\ 
\frac{-q^{2/3}}{r} &  & s=\frac{3}{4} \\ 
&  &  \\ 
\frac{\left( 2s-1\right) \left( qr^{-2s}\right) ^{\frac{1-s}{s\left(
2s-1\right) }}}{2\left( s-1\right) } &  & \text{otherwise}%
\end{array}%
\right. ,  \label{h(r)}
\end{equation}%
where $q$ is an integration constant related to the electric charge and $l$
is an arbitrary constant with length dimension, which comes from the fact
that the logarithmic arguments should be dimensionless.

It is notable that the electromagnetic field tensor is given by 
\begin{equation}
F_{tr}=E(r)=\left\{ 
\begin{array}{ccc}
\frac{q}{lr} &  & s=1 \\ 
&  &  \\ 
\frac{q^{2/3}}{r^{2}} &  & s=\frac{3}{4} \\ 
&  &  \\ 
\left( qr^{\frac{-s}{1-s}}\right) ^{\frac{1-s}{s\left( 2s-1\right) }} &  & 
\text{otherwise}%
\end{array}%
\right. .  \label{E(r)}
\end{equation}

Worth mentioning that the electromagnetic gauge potential (\ref{h(r)}) and
the electromagnetic field (\ref{E(r)}), should be finite at infinity. These
constraints impose the following restriction on the nonlinearity parameter ($%
s$) as 
\begin{equation}
\frac{1}{2}<s\leq 1.
\end{equation}

By considering the above constraint, hereafter, the expression \emph{%
"otherwise"} belongs to the range $\left\{ \frac{1}{2}<s<\frac{3}{4} \right\}
\cup \left\{ \frac{3}{4}<s<1\right\} $, or $s\in \left( \frac{1}{2},\frac{3}{%
4}\right) \cup \left( \frac{3}{4},1\right) $.

Now, we would like to obtain exact solutions. For this purpose, by
employing Eq. (\ref{massiveTerm1}), the metric ansatz (\ref{f11}), and the
metric (\ref{metric}), we obtain the massive terms in the following forms
\begin{eqnarray}
\chi _{tt} &=&\frac{c\varepsilon g(r)}{2r},  \notag \\
&&  \notag \\
\chi _{rr} &=&\frac{-c\varepsilon }{2rg(r)},  \label{Xx} \\
&&  \notag \\
\chi _{\varphi \varphi } &=&0,  \notag
\end{eqnarray}%
where $\chi _{tt}$, $\chi _{rr}$, and $\chi _{\varphi\varphi }$ are
corresponding to $tt$, $rr$, and $\varphi \varphi $ components of Eq. (\ref%
{massiveTerm}), respectively. Now, we calculate the non-zero components of $%
G_{\mu \nu}+\Lambda g_{\mu \nu }=0$, by using the metric (\ref{metric}),
which are  
\begin{eqnarray}
G_{tt}+\Lambda g_{tt} &=&-\frac{g(r)\left[ g^{\prime }(r)+2\Lambda r\right] 
}{2r},  \notag \\
&&  \notag \\
G_{rr}+\Lambda g_{rr} &=&\frac{g^{\prime }(r)+2\Lambda }{2rg\left( r\right) }%
,  \label{Einstein1} \\
&&  \notag \\
G_{\varphi \varphi }+\Lambda g_{\varphi \varphi } &=&\frac{r^{2}}{2}%
g^{\prime \prime }(r)+\Lambda r^{2},  \notag
\end{eqnarray}%
by considering Eqs. (\ref{Xx}) and (\ref{Einstein1}), we can obtain the
non-zero components of Eq. (\ref{Field equation}), as 
\begin{eqnarray}
E_{tt} &=&E_{rr}=rg^{\prime }(r)+2\Lambda r^{2}-m_{g}^{2}c{\small %
\varepsilon }r+2^{s}\left( 2s-1\right) \Psi _{2}=0,  \label{eqENMax1} \\
&&  \notag \\
E_{\varphi \varphi } &=&\frac{r^{2}g^{\prime \prime }(r)}{2}+\Lambda
r^{2}-2^{s-1}\Psi _{2}=0,  \label{eqENMax2}
\end{eqnarray}%
where $\Psi _{2}$ is 
\begin{equation}
\Psi _{2}=\left\{ 
\begin{array}{ccc}
\frac{q^{2}}{l^{2}} &  & s=1 \\ 
&  &  \\ 
\frac{q}{r} &  & s=\frac{3}{4} \\ 
&  &  \\ 
\left( \frac{q}{r}\right) ^{\frac{2\left( 1-s\right) }{2s-1}} &  & s\in
\left( \frac{1}{2},\frac{3}{4}\right) \cup \left( \frac{3}{4},1\right)%
\end{array}%
\right. ,
\end{equation}%
which $E_{tt}$, $E_{rr}$ and $E_{\varphi \varphi }$ are corresponding to $tt$%
, $rr$ and $\varphi \varphi $ components of Eq. (\ref{Field equation}),
respectively. After some manipulations, one can obtain the following metric
function 
\begin{equation}
g(r)=-m_{0}-\Lambda r^{2}+m_{g}^{2}c{\small \varepsilon }r+\left\{ 
\begin{array}{ccc}
\frac{-2q^{2}}{l^{2}}\ln \left( \frac{r}{l}\right) &  & s=1 \\ 
&  &  \\ 
\frac{q}{2^{1/4}r} &  & s=\frac{3}{4} \\ 
&  &  \\ 
\frac{2^{s-1}\left( 2s-1\right) ^{2}\left( \frac{q}{r}\right) ^{\frac{%
2\left( 1-s\right) }{2s-1}}}{\left( 1-s\right) } &  & s\in \left( \frac{1}{2}%
,\frac{3}{4}\right) \cup \left( \frac{3}{4},1\right)%
\end{array}%
\right. ,  \label{f(r)ENMax}
\end{equation}%
where $m_{0}$ is an integration constant related to the black hole's total
mass. We should note that the obtained metric function simultaneously
satisfies all field equation components (\ref{Field equation}).

To examine the geometrical structure of these solutions, first, we look for
essential singularity(ies) by calculating the Ricci and Kretschmann scalars.
We obtain these scalars in the following forms 
\begin{eqnarray}
R &=&6\Lambda -\frac{2m_{g}^{2}c{\small \varepsilon }}{r}+\left\{ 
\begin{array}{ccc}
\frac{2q^{2}}{l^{2}r^{2}} &  & s=1 \\ 
&  &  \\ 
0 &  & s=\frac{3}{4} \\ 
&  &  \\ 
\frac{2^{s}\left( 4s-3\right) }{r^{2}}\left( \frac{q}{r}\right) ^{\frac{%
2\left( 1-s\right) }{2s-1}} &  & s\in \left( \frac{1}{2},\frac{3}{4}\right)
\cup \left( \frac{3}{4},1\right)%
\end{array}%
\right. ,  \label{R} \\
&&  \notag \\
R_{\alpha \beta \gamma \delta }R^{\alpha \beta \gamma \delta } &=&12\Lambda
^{2}-\frac{8\Lambda m_{g}^{2}c{\small \varepsilon }}{r}+\frac{2m_{g}^{4}c^{2}%
{\small \varepsilon }^{2}}{r^{2}}  \notag \\
&&  \notag \\
&&+\left\{ 
\begin{array}{ccc}
\frac{8\Lambda q^{2}}{l^{2}r^{2}}-\frac{8q^{2}m_{g}^{2}c\varepsilon }{%
l^{2}r^{3}}+\frac{12q^{4}}{l^{4}r^{4}} &  & s=1 \\ 
&  &  \\ 
-\frac{2^{7/4}qm_{g}^{2}c\varepsilon }{r^{4}}+\frac{3\sqrt{2}q^{2}}{r^{6}} & 
& s=\frac{3}{4} \\ 
&  &  \\ 
\frac{2^{s+2}\left( \frac{q}{r}\right) ^{\frac{4\left( 1-s\right) }{2s-1}}}{%
r^{4}}\left( \frac{\Lambda \left( 4s-3\right) r^{2}-\left( 2s-1\right)
m_{g}^{2}c\varepsilon r}{\left( \frac{q}{r}\right) ^{\frac{2\left(
1-s\right) }{2s-1}}}+2^{s+1}\left( s^{2}-s+\frac{3}{8}\right) \right) &  & 
s\in \left( \frac{1}{2},\frac{3}{4}\right) \cup \left( \frac{3}{4},1\right)%
\end{array}%
\right. .  \label{K}
\end{eqnarray}

It is evident that the Ricci and Kretschmann scalars diverge at the origin as%
\begin{eqnarray}
\underset{r\rightarrow 0}{\lim }R &=&\infty ,  \notag \\
&& \\
\underset{r\rightarrow 0}{\lim }R_{\alpha \beta \gamma \delta }R^{\alpha
\beta \gamma \delta } &=&\infty ,  \notag
\end{eqnarray}%
so, there is a curvature singularity at $r=0$.

Also, for large values of radial coordinate, $r\longrightarrow \infty $, the
Ricci and Kretschmann scalars are%
\begin{eqnarray}
\underset{r\rightarrow \infty }{\lim }R &=&6\Lambda ,  \notag \\
&& \\
\underset{r\rightarrow \infty }{\lim }R_{\alpha \beta \gamma \delta
}R^{\alpha \beta \gamma \delta } &=&12\Lambda ^{2},  \notag
\end{eqnarray}%
in which confirm that the solution's asymptotical behavior is dS for $%
\Lambda >0$ and AdS for $\Lambda <0$. However, there is a study on
three-dimensional black holes that shows that classical black holes do not
exist in three-dimensional dS spacetime \cite{Emparan}. Therefore, we
consider the AdS case in our work.

\begin{figure}[tbh]
\centering
\includegraphics[width=0.3\textwidth]{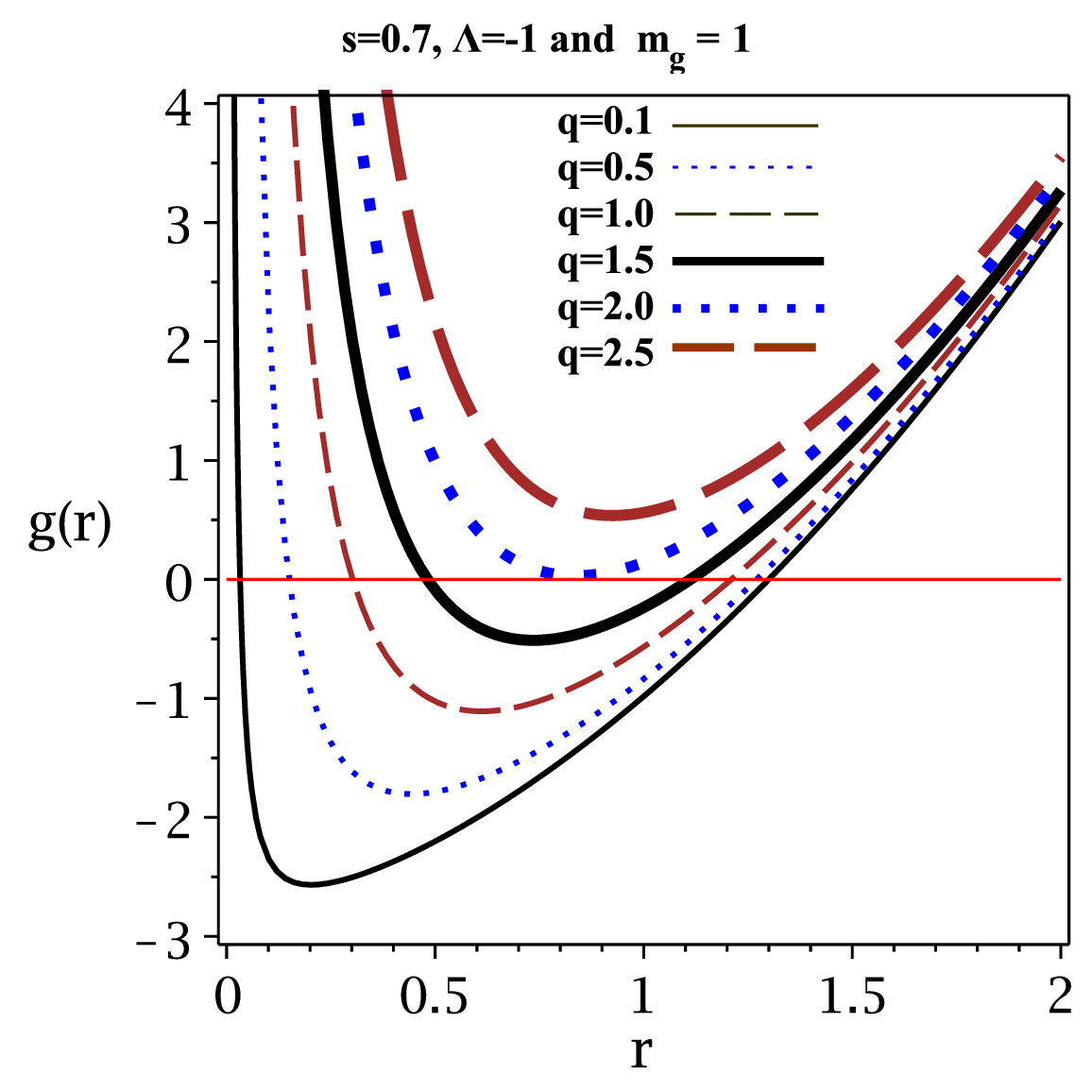} %
\includegraphics[width=0.3\textwidth]{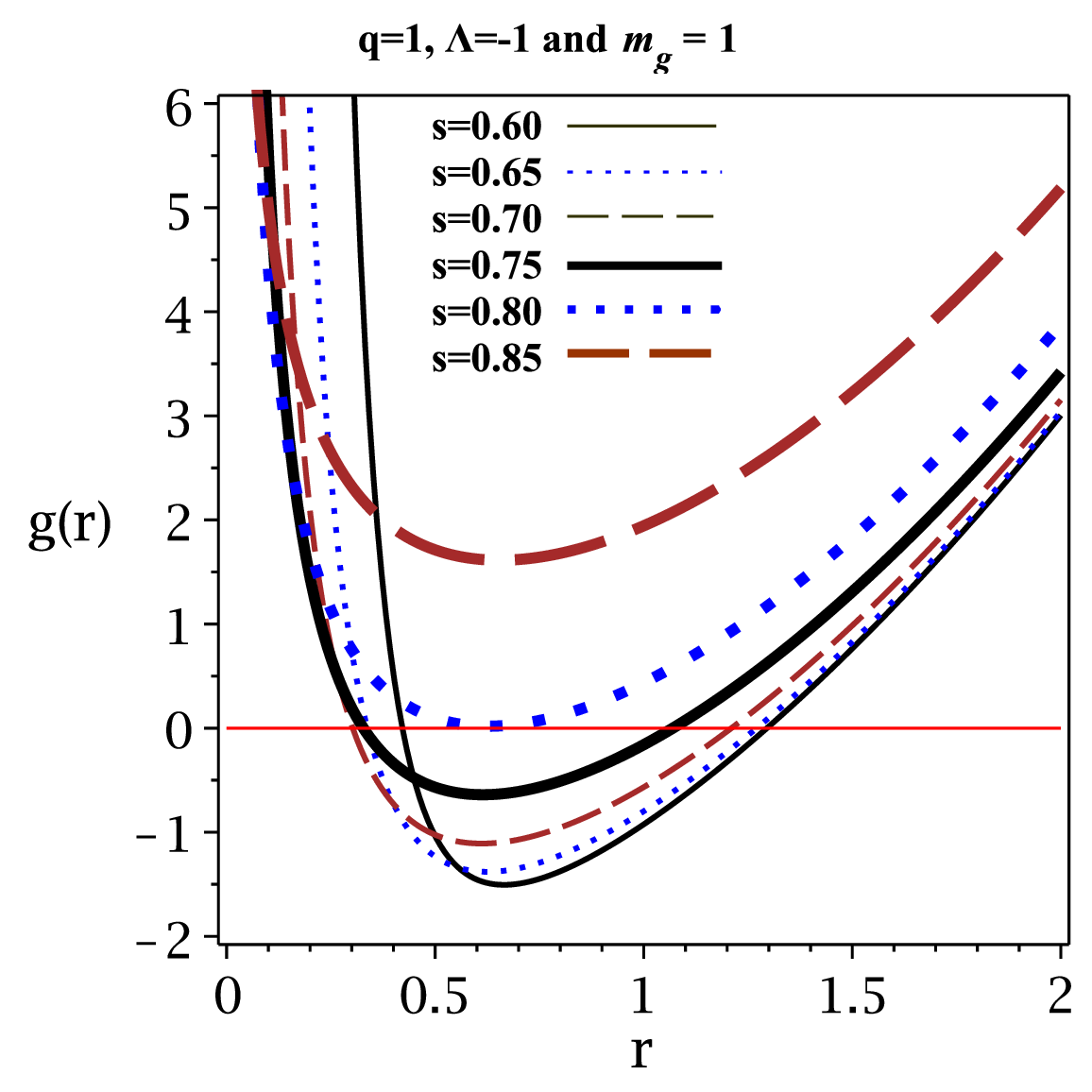} %
\includegraphics[width=0.3\textwidth]{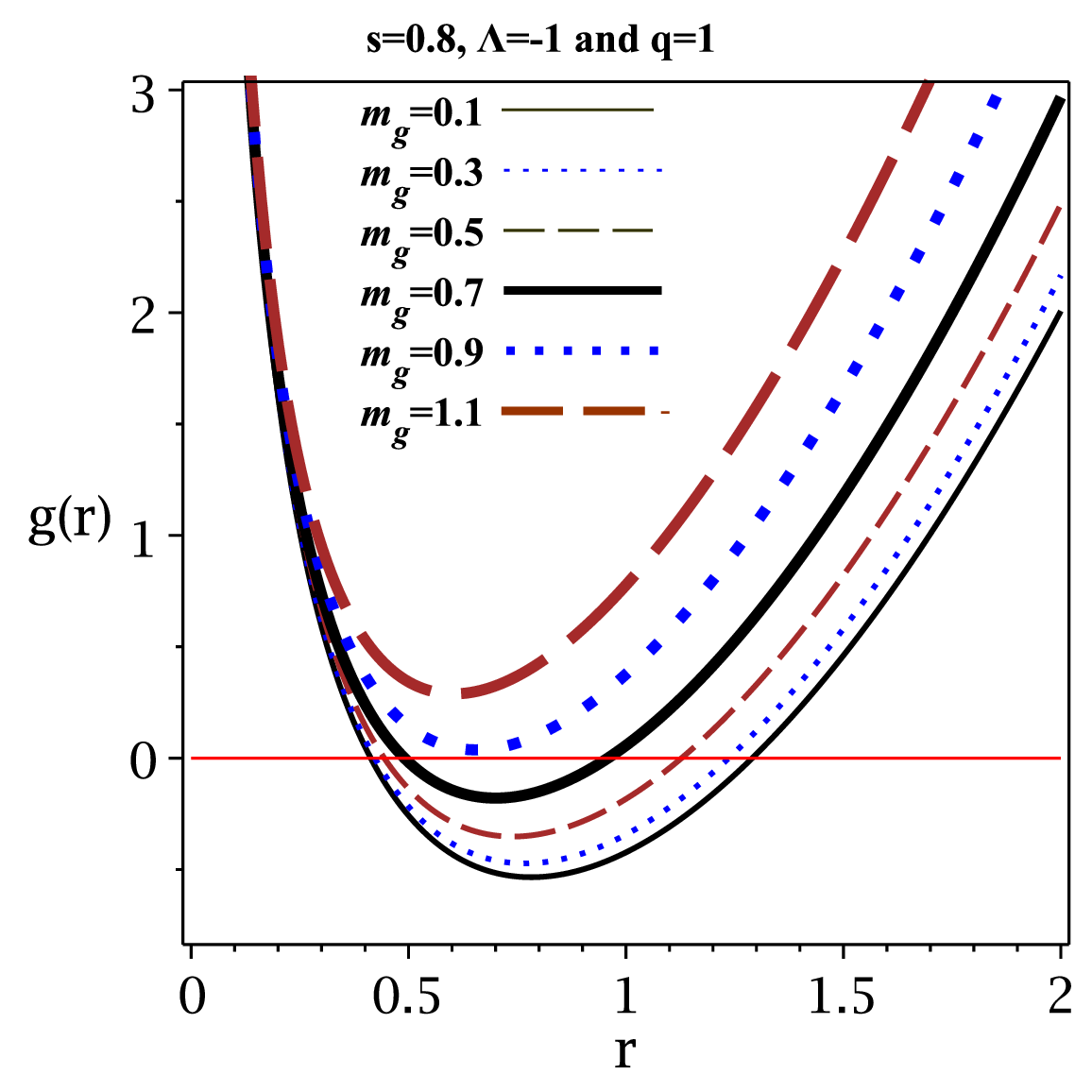} \newline
\caption{$g(r)$ versus $r$ for ${\protect\small \protect\varepsilon =1}$, $%
c=1$ and $m_{0}=3$.}
\label{Fig1}
\end{figure}

In order to study the effects of massive gravitons ($m_{g}$), the parameter
of PM theory ($s$), electrical charge ($q$), and the cosmological constant ($%
\Lambda $), one can investigate the metric function (\ref{f(r)ENMax}).
Regarding various terms of $g(r)$, it is worthwhile to mention that $q$-term
(the fourth term in Eq. (\ref{f(r)ENMax})) is dominant near the origin ($%
r\rightarrow 0$). Therefore, one can conclude that the singularity is
timelike. In addition, for large distance ($r\rightarrow \infty $), $\Lambda$%
-term (the second term in Eq. (\ref{f(r)ENMax})) is dominant, which confirms
that the solutions can be asymptotically AdS. As one can see, the behavior
of $g(r)$ is highly sensitive to the massive graviton, the parameter of PM
theory, the electrical charge, and the cosmological constant (see Fig. \ref%
{Fig1}, for more details). It is evident that for specific values of
different parameters, the metric function could have two roots, one extreme
root or no root (see up panels in Fig. \ref{Fig1}). In addition, the
three-dimensional solutions are covered by an event horizon, which confirms
the existence of black hole solutions in three-dimensional spacetime. 

\section{Thermodynamics}

\label{SecIV} Now, we intend to calculate the conserved and thermodynamic
quantities of these black hole solutions and show thermodynamic quantities of this theory satisfy the first law of thermodynamics. For studying the thermodynamic properties of three-dimensional black
holes in massive-power-Maxwell theory, it is necessary to express the mass ($%
m_{0}$) in terms of the radius of the event horizon $r_{e}$, electrical
charge ($q$), and the cosmological constant ($\Lambda $), mass of graviton ($%
m_{g}$), and the parameters of massive gravity ($c$, and $\varepsilon $).
Equating $g(r)=0$ (Eq. (\ref{f(r)ENMax})), we obtain
\begin{equation}
m_{0}=-\Lambda r_{e}^{2}+m_{g}^{2}c{\small \varepsilon }r_{e}-\left\{ 
\begin{array}{ccc}
\frac{2q^{2}\ln \left( \frac{r_{e}}{l}\right) }{l^{2}}, &  & s=1 \\ 
&  &  \\ 
\frac{-2^{3/4}q}{2r_{e}}, &  & s=\frac{3}{4} \\ 
&  &  \\ 
\frac{2^{s-1}\left( 2s-1\right) ^{2}}{\left( s-1\right) }\left( \frac{q}{%
r_{e}}\right) ^{\frac{2\left( 1-s\right) }{2s-1}}, &  & s\in \left( \frac{1}{%
2},\frac{3}{4}\right) \cup \left( \frac{3}{4},1\right)%
\end{array}%
\right. .
\end{equation}

The Hawking temperature is defined as
\begin{equation}
T=\frac{\kappa }{2\pi },  \label{T0}
\end{equation}%
where $\kappa $ denotes the surface gravity by
\begin{equation}
\kappa =\sqrt{\frac{-1}{2}\left( \nabla _{\mu }\chi _{\nu }\right) \left(
\nabla ^{\mu }\chi ^{\nu }\right) },  \label{k}
\end{equation}%
in which $\chi $ is Killing vector. For the mentioned spacetime, the
Killing vector is $\chi =\partial _{t}$. Therefore, by considering the
metric (\ref{metric}), we can obtain the surface gravity in the following
form  
\begin{equation}
\kappa =\frac{1}{2}\left. \frac{dg\left( r\right) }{dr}\right\vert
_{r=r_{e}},  \label{k1}
\end{equation}%
by considering the metric function $g(r)$ (\ref{f(r)ENMax}), the
surface gravity (\ref{k1}), and the Hawking temperature (\ref{T0}), we find 
\begin{equation}
T=-\frac{\Lambda r_{e}}{2\pi }+\frac{m_{g}^{2}c{\small \varepsilon }}{4\pi }%
-\left\{ 
\begin{array}{ccc}
\frac{q^{2}}{2\pi l^{2}r_{e}} &  & s=1 \\ 
&  &  \\ 
\frac{q}{2^{9/4}\pi r_{e}^{2}} &  & s=\frac{3}{4} \\ 
&  &  \\ 
\frac{2^{s-2}\left( 2s-1\right) }{\pi r_{e}}\left( \frac{q}{r_{e}}\right) ^{%
\frac{2\left( 1-s\right) }{2s-1}} &  & s\in \left( \frac{1}{2},\frac{3}{4}%
\right) \cup \left( \frac{3}{4},1\right)%
\end{array}%
\right. .  \label{TotalTT}
\end{equation}

According to Gauss's law, the electric charge, $Q$, can be found by
calculating the flux of the electric field at infinity. For example,
we apply Gauss's law for Maxwell theory ($s=1$) which yields 
\begin{equation}
Q_{Maxwell}=\left. \frac{1}{4\pi }\int_{0}^{2\pi }F_{tr_{Maxwell}}\sqrt{%
g_{\varphi \varphi }}d\varphi \right\vert _{r=r_{e}}=\frac{q}{4\pi l}%
\int_{0}^{2\pi }d\varphi =\frac{q}{2l}
\end{equation}%
where $F_{tr_{Maxwell}}=\frac{q}{lr}$ is the Maxwell electromagnetic
field (\ref{E(r)}), and $g_{\varphi \varphi }=r^{2}$, is $\varphi \varphi $
component of the metric tensor ($g_{\mu\nu }$). In addition, the electric
charge of the PM theory is obtained in Ref. \cite{Hendi2016}, which is 
\begin{equation}
Q=\left\{ 
\begin{array}{ccc}
\frac{3q^{1/3}}{2^{13/4}} &  & s=\frac{3}{4} \\ 
&  &  \\ 
2^{s-2}sq^{\frac{1-s}{s}} &  & s\in \left( \frac{1}{2},\frac{3}{4}\right)
\cup \left( \frac{3}{4},1\right)%
\end{array}%
\right. .  \label{TotalQ}
\end{equation}

To obtain the total mass of the solutions, we follow the obtained
result in Ref. \cite{VeghBHI}. Indeed, the ADM mass can be obtained through
the Hamiltonian approach which is given by $M=\frac{\left(d-2\right) \omega
_{d}}{16\pi }m_{0}$ \cite{VeghBHI}, which $d$, and $\omega _{d}$ are related
to the dimension of spacetime, and unit volume in $d$-dimension, respectively. For
three-dimensional spacetime, we have $d=3$, and $\omega
_{d=3}=\int_{0}^{2\pi}d\varphi =2\pi $. So, the total mass of the solutions
is given by
\begin{equation}
M=\frac{m_{0}}{8},  \label{TotalM}
\end{equation}%
in which by evaluating metric function on the horizon ($g\left(
r=r_{e}\right) =0$), we obtain 
\begin{equation}
M=\frac{-\Lambda r_{e}^{2}+m_{g}^{2}c{\small \varepsilon }r_{e}}{8}+\left\{ 
\begin{array}{ccc}
-\frac{q^{2}\ln \left( \frac{r_{e}}{l}\right) }{4l^{2}} &  & s=1 \\ 
&  &  \\ 
\frac{q}{2^{13/4}r_{e}} &  & s=\frac{3}{4} \\ 
&  &  \\ 
\frac{2^{s-4}\left( 2s-1\right) ^{2}\left( \frac{q}{re}\right) ^{\frac{%
2\left( 1-s\right) }{2s-1}}}{\left( 1-s\right) } &  & s\in \left( \frac{1}{2}%
,\frac{3}{4}\right) \cup \left( \frac{3}{4},1\right)%
\end{array}%
\right. .  \label{TotalM1}
\end{equation}

To obtain the entropy of black holes in the presence of
Einstein-massive gravity, one can use the area law proposed by Hawking and
Bekenstein in the following form 
\begin{equation}
S=\frac{A}{4},  \label{SS1}
\end{equation}%
where $A$ is the horizon area and for three-dimensional spacetime is
defined 
\begin{equation}
A=\int_{0}^{2\pi }\left. \sqrt{g_{\varphi \varphi }}d\varphi \right\vert
_{r=r_{e}}=2\pi r_{e},  \label{SS2}
\end{equation}%
by replacing the horizon area (\ref{SS2}) within Eq. (\ref{SS1}), we
obtain the entropy as
\begin{equation}
S=\frac{\pi }{2}r_{e}.  \label{TotalS}
\end{equation}

The electric potential, $U$, is defined through the gauge potential in the
following form 
\begin{equation}
U=A_{\mu }\chi ^{\mu }\left\vert _{r\rightarrow reference}\right. -A_{\mu
}\chi ^{\mu }\left\vert _{r\rightarrow r_{e}}\right. =\left\{ 
\begin{array}{ccc}
-\frac{q}{l}\ln \left( \frac{r_{e}}{l}\right) &  & s=1 \\ 
&  &  \\ 
\frac{q^{2/3}}{r_{e}} &  & s=\frac{3}{4} \\ 
&  &  \\ 
\frac{\left( 2s-1\right) \left( qr_{e}^{-2s}\right) ^{\frac{1-s}{s\left(
2s-1\right) }}}{2\left( 1-s\right) } &  & s\in \left( \frac{1}{2},\frac{3}{4}%
\right) \cup \left( \frac{3}{4},1\right)%
\end{array}%
\right. ,  \label{TotalU}
\end{equation}%
where $A_{\mu }=h\left( r\right) \delta _{\mu }^{t}$ is obtained in
Eq. (\ref{h(r)}).

Having conserved and thermodynamic quantities at hand, we are in a
position to check whether thermodynamic quantities satisfy the first law of
thermodynamics. It is easy to show that by using thermodynamic quantities
such as electric charge (\ref{TotalQ}), mass (\ref{TotalM}), and entropy (%
\ref{TotalS}), with the first law of black hole thermodynamics 
\begin{equation}
dM=TdS+UdQ,
\end{equation}%
one can define the intensive parameters conjugate to $S$ and $Q$. These
quantities are the temperature and the electric potential 
\begin{equation}
T=\left( \frac{\partial M}{\partial S}\right) _{Q}~~~\&~~~U=\left( \frac{%
\partial M}{\partial Q}\right) _{S},  \label{TU}
\end{equation}%
which are the same as those calculated for the temperature (\ref{TotalTT})
and the electric potential (\ref{TotalU}). Evidently, the obtained
thermodynamic quantities could satisfy the first law of thermodynamics.

\subsection{Thermal stability in the canonical ensemble}

Here, we study thermal stability criteria and the effects of different
parameters on them. The stability conditions in canonical ensemble are based
on the sign of the heat capacity. This change of sign could happen when heat
capacity meets root(s) or divergency(ies). The root of heat capacity (or
temperature) indicates a bound point, which separates physical solutions
(positive temperature) from non-physical ones (negative temperature). The
heat capacity divergencies (the roots of the denominator of heat capacity)
represent phase transition points. The negativity of heat capacity
represents unstable solutions that may undergo a phase transition and
acquire a stable state. To get a better picture and enrich our study's
results, we investigate temperature and heat capacity simultaneously.

The heat capacity is given by the following traditional relation 
\begin{equation}
C_{Q}=\frac{T}{\left( \frac{\partial ^{2}M}{\partial S^{2}}\right) _{Q}}=%
\frac{T}{{\left( \frac{\partial T}{\partial S}\right) _{Q}}}.  \label{CQ}
\end{equation}

Considering Eqs. (\ref{TotalTT}) and (\ref{TotalS}), it is a matter of the
calculation to show that 
\begin{equation}
C_{Q}=\frac{\left( 2\Lambda r_{e}^{2}-m_{g}^{2}c{\small \varepsilon }%
r_{e}+2^{s}\left( 2s-1\right) \Psi _{3}\right) \pi r_{e}}{4\Lambda
r_{e}^{2}-2^{s+1}\Psi _{3}},  \label{CQMax}
\end{equation}%
where 
\begin{equation}
\Psi _{3}=\left\{ 
\begin{array}{ccc}
\frac{q^{2}}{l^{2}} &  & s=1 \\ 
&  &  \\ 
\frac{q}{r_{e}} &  & s=\frac{3}{4} \\ 
&  &  \\ 
\left( \frac{q}{r_{e}}\right) ^{\frac{2\left( 1-s\right) }{2s-1}} &  & s\in
\left( \frac{1}{2},\frac{3}{4}\right) \cup \left( \frac{3}{4},1\right)%
\end{array}%
\right. .
\end{equation}

Notably, the obtained temperature (\ref{TotalTT}) and the heat capacity (\ref%
{CQMax}) include three terms: cosmological constant, electric charge, and
massive terms.

\begin{figure}[tbh]
\centering
\includegraphics[width=0.3\textwidth]{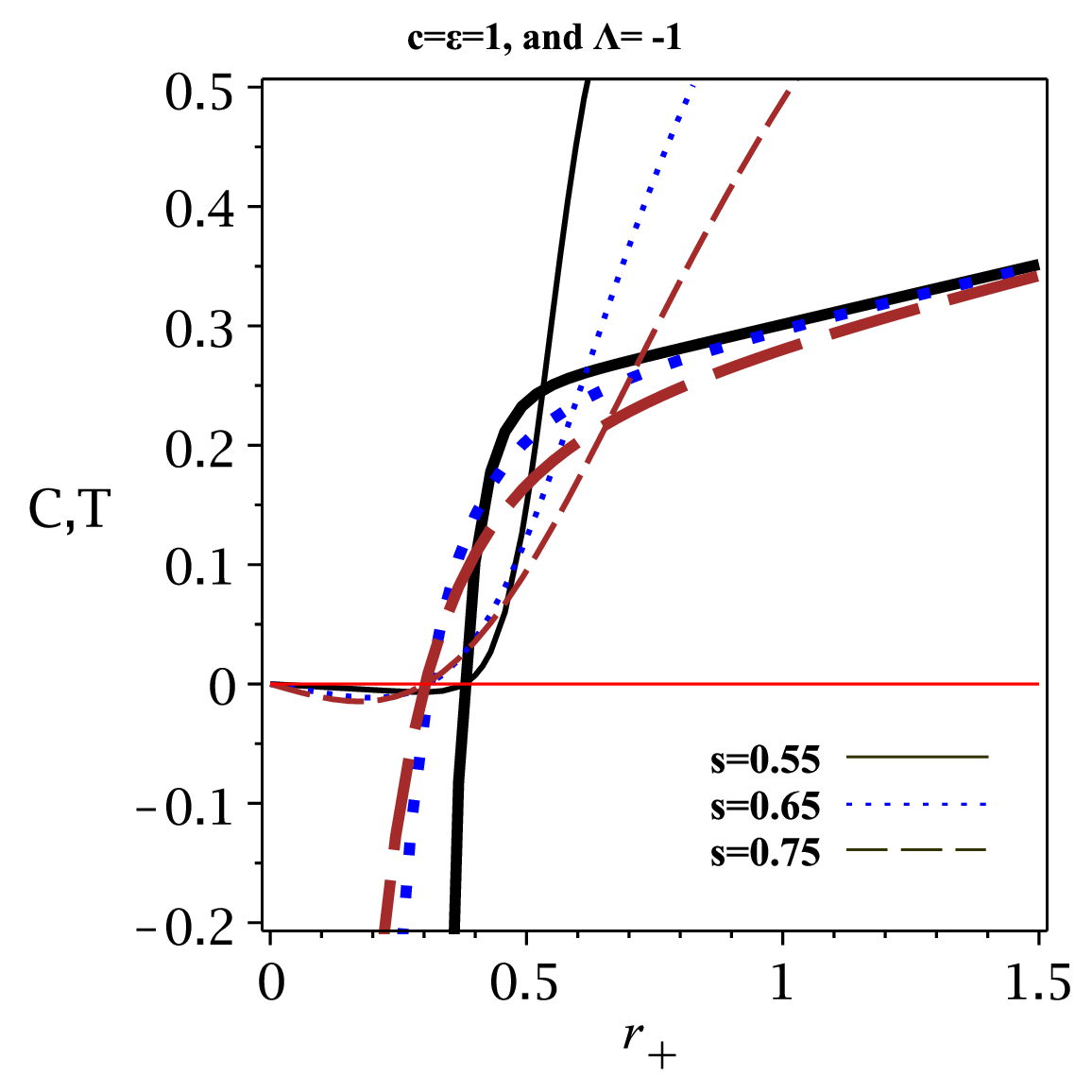} \includegraphics[width=0.3%
\textwidth]{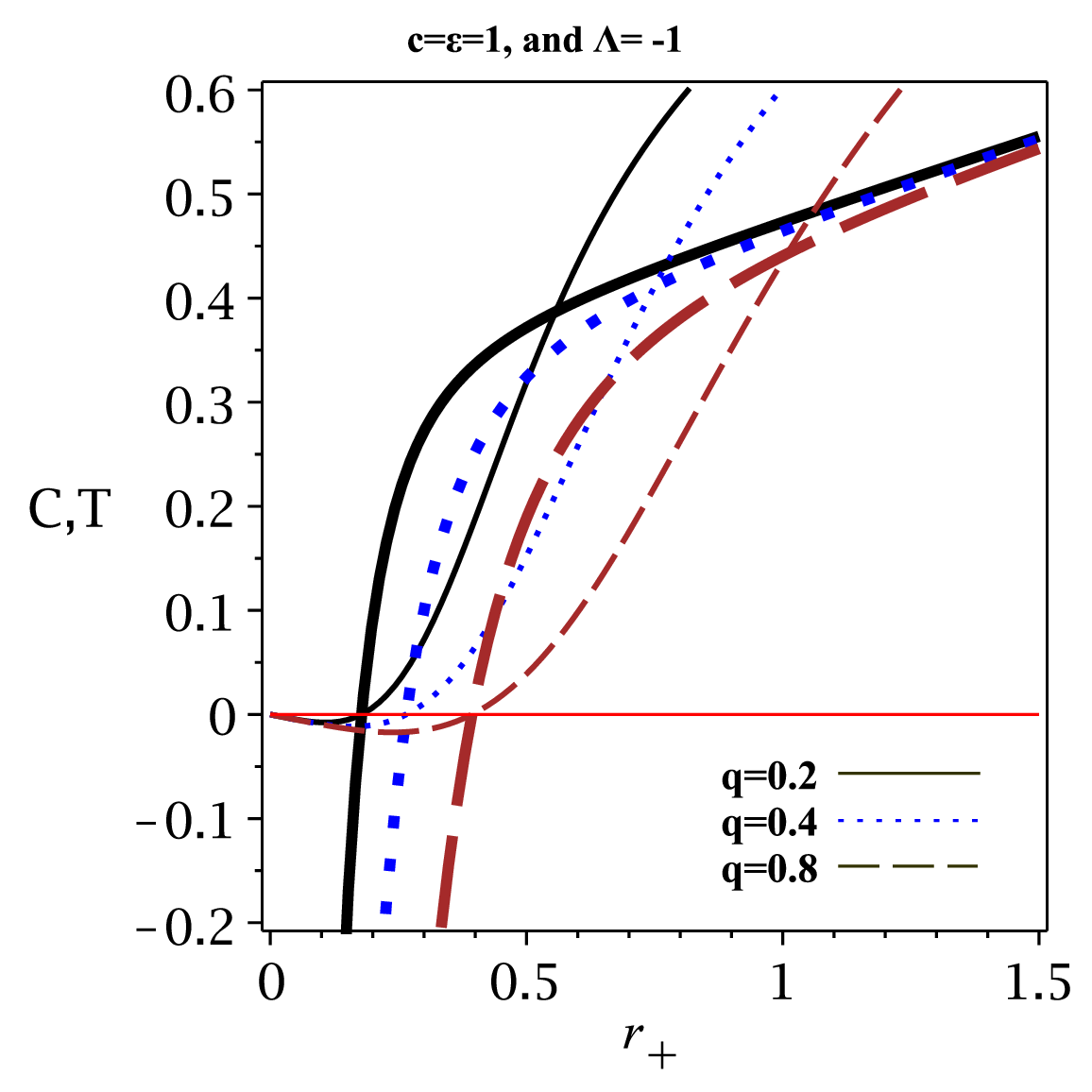} \includegraphics[width=0.3\textwidth]{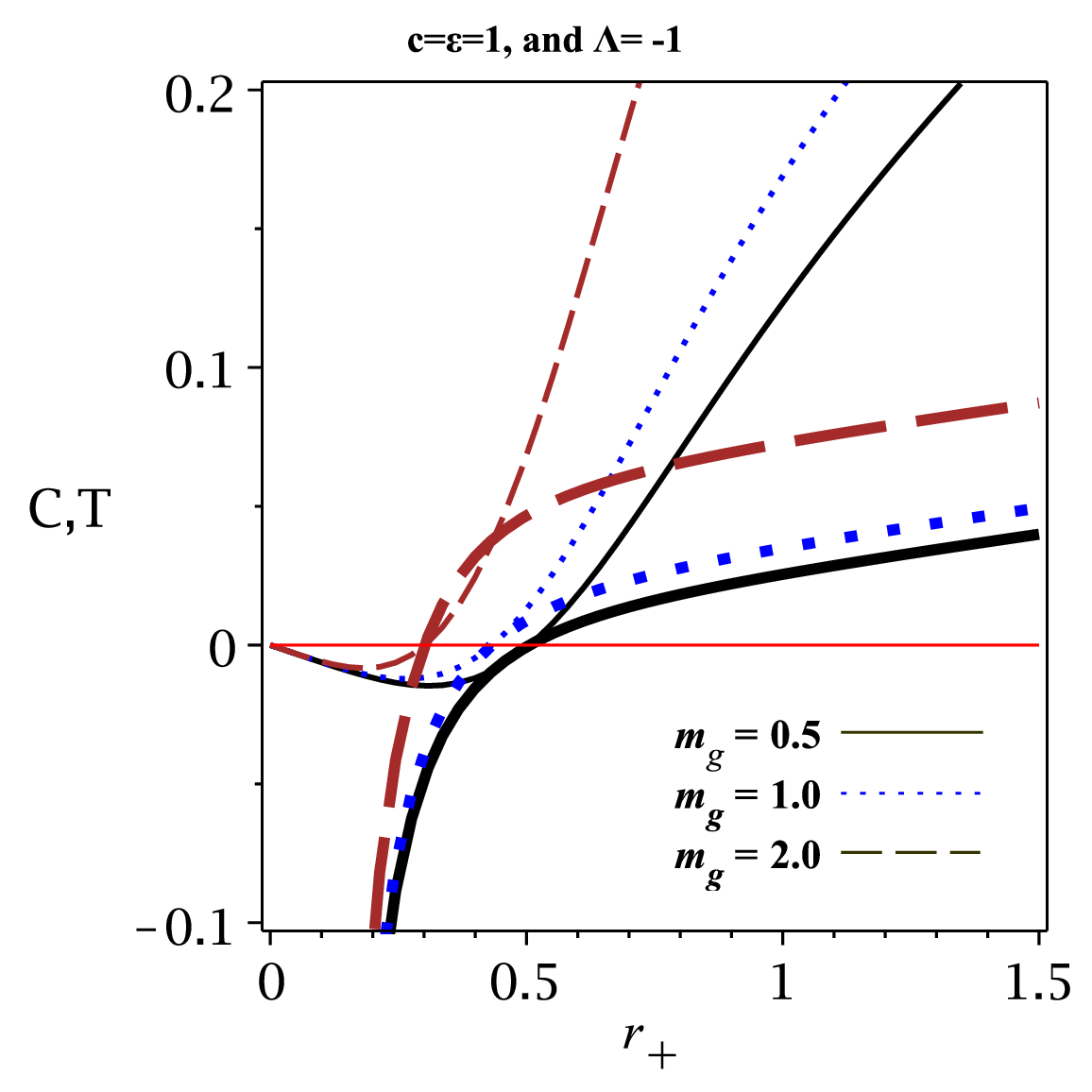} 
\newline
\caption{$T$ (Bold lines) and $C_{Q}$ (thin lines) versus $r_{+}$ for
different $s$ (left panel), different $q$ (middle panel) and different $%
m_{g} $ (right panel).}
\label{Fig3}
\end{figure}

According to the obtained heat capacity for cases $s=1$ and $s=\frac{3}{4}$,
we are in a position to find the exact bound and phase transition points of
AdS black holes. Solving the numerator and denominator of the heat capacity
concerning the horizon's radius leads to the following solutions for bound
points ($r_{b}$) and phase transition points ($r_{p}$), respectively, 
\begin{eqnarray}
r_{b} &=&\left\{ 
\begin{array}{ccc}
\frac{m_{g}^{2}c\varepsilon -\sqrt{m_{g}^{4}c^{2}\varepsilon ^{2}-\frac{%
16\Lambda q^{2}}{l^{2}}}}{4\Lambda } &  & s=1 \\ 
&  &  \\ 
\frac{\Psi _{4}+\frac{m_{g}^{4}c^{2}\varepsilon ^{2}}{\Psi _{4}}%
+m_{g}^{2}c\varepsilon }{4\Lambda } &  & s=\frac{3}{4}%
\end{array}%
\right. ,  \label{rb} \\
&&  \notag \\
r_{p} &=&\left\{ 
\begin{array}{ccc}
\pm \frac{q}{l\sqrt{\Lambda }} &  & s=1 \\ 
&  &  \\ 
\frac{\left( 2^{11/4}q\Lambda ^{2}\right) ^{1/3}}{2\Lambda } &  & s=\frac{3}{%
4}%
\end{array}%
\right. ,  \label{rp}
\end{eqnarray}%
where $\Psi _{4}=\left( -3^{3}2^{3/4}q\Lambda ^{2}+m_{g}^{6}c^{3}\varepsilon
^{3}+3^{3/2}\Lambda \sqrt{2^{3/4}q\left[ 3^{3}2^{3/4}q\Lambda
^{2}-2m_{g}^{6}c^{3}\varepsilon ^{3}\right] }\right) ^{1/3}$. Interestingly,
phase transition cannot exist for AdS black holes, and it is independent of
the massive term (Eq. (\ref{rp})). Another interesting result is related to
the effects of massive term and the electric charge on the bound points of
AdS black holes for cases $s=1$ and $s=\frac{3}{4}$. Considering Eq. (\ref%
{rb}), it is clear that the bound points shift to the larger (smaller)
horizon radius by increasing the electric charge (massive term). Indeed, the
physical area decreases for higher charged AdS black holes (see the middle
and right panels in Fig. \ref{Fig3}). For different values of $s$, the
physical area increases by increasing $s$ (see the left panel in Fig. \ref%
{Fig3}).

\section{Optical features}

In this section, we carefully study the optical features of AdS black holes
in three-dimensional PM-massive theory, such as the photon orbit, the energy emission rate, and the deflection angle. Considering these
optical quantities, we investigate the influence of parameters of the theory
on the black hole solutions.

\subsection{Null geodesics and photon orbit}

\label{SecVA} Here, we would like to obtain the radius of the photon orbit
and critical impact parameter for the corresponding black hole and show how they are
affected by solution parameters. To this purpose, we employ the
Hamilton-Jacobi method for a photon in the black hole spacetime as \cite%
{Carter1a,Decanini1a} 
\begin{equation}
\frac{\partial \mathcal{S}}{\partial \sigma }+H=0,
\end{equation}%
where $\mathcal{S}$ and $\sigma $ are the Jacobi action and affine parameter
along the geodesics, respectively. The geodesic motion of a massless photon
in the static spherically symmetric spacetime can be controlled by the
following Hamiltonian 
\begin{equation}
H=\frac{1}{2}g^{ij}p_{i}p_{j}=0.  \label{EqHamiltonian}
\end{equation}

Taking into account Eq. (\ref{metric}), the above equation can be written as 
\begin{equation}
\frac{1}{2}\left[ -\frac{p_{t}^{2}}{g(r)}+g(r)p_{r}^{2}+\frac{p_{\varphi
}^{2}}{r^{2}}\right] =0,  \label{EqNHa}
\end{equation}%
from which we deduce 
\begin{equation}
\dot{p_{t}}=-\frac{\partial H}{\partial t}=0,~~~~\&~~~~\dot{p_{\varphi }}=-%
\frac{\partial H}{\partial \varphi }=0,  \label{Eqpt1}
\end{equation}

This shows that the Hamiltonian is independent of the coordinates $t$ and $%
\varphi $. So, one can consider $p_{t} $ and $p_{\varphi} $ as constants of
motion. We define $- p_{t} \equiv E$ and $p_{\varphi} \equiv L $ where $E $
and $L $ are, respectively, the energy and angular momentum of the photon.

Using the Hamiltonian formalism, the equations of motion are given by 
\begin{equation}
\dot{t}=\frac{\partial H}{\partial p_{t}}=-\frac{p_{t}}{g(r)},~~~\&~~~\dot{r}%
=\frac{\partial H}{\partial p_{r}}=p_{r}g(r),~~~\&~~~\dot{\varphi}=\frac{%
\partial H}{\partial p_{\varphi }}=\frac{p_{\varphi }}{r^{2}},  \label{Eqem}
\end{equation}%
where $p_{r}$ is the radial momentum and the overdot denotes a derivative
with respect to the affine parameter $\sigma $. These equations and two
conserved quantities provide a complete description of the dynamics by
taking into account the orbital equation of motion as follows 
\begin{equation}
\dot{r}^{2}+V_{\mathrm{eff}}(r)=0,  \label{EqVef1}
\end{equation}%
where $V_{\mathrm{eff}}$ is the effective potential of the photon, given by 
\begin{equation}
V_{\mathrm{eff}}(r)=g(r)\left[ \frac{L^{2}}{r^{2}}-\frac{E^{2}}{g(r)}\right]
.  \label{Eqpotential}
\end{equation}

It should be noted that the photon orbits are circular and unstable,
associated with the maximum value of the effective potential. The following
conditions can obtain such a maximum

\begin{equation}
V_{\mathrm{eff}}(r_{ph})=0,~~~\&~~~V_{\mathrm{eff}}^{\prime
}(r_{ph})=0,~~~\&~~~V_{\mathrm{eff}}^{\prime \prime }(r_{ph})<0,
\label{EqVeff1}
\end{equation}%
where the first two conditions determine the critical angular momentum of
the photon orbit $(L_{p})$ and the photon orbit radius $(r_{ph})$,
respectively, also, the third condition ensures that the photon orbits are
unstable. The impact parameter which is defined by the ratio of
angular momentum and energy of the photon is obtained as 
\begin{equation}
b=\frac{L}{E},
\end{equation}
and the critical impact parameter is defined as $b_{c}=\frac{L_{p}}{E} $. An
ingoing photon from infinity with $b<b_{c}$ falls into the black hole
without reaching the observer, whereas it bounces back if $b>b_{c}$ and can
be observed by the observer located at infinity. An interesting phenomenon
is related to the critical impact parameter for which $max(V_{eff}) = 0$. In
this case, the ingoing photon loses both its radial velocity and
acceleration at $r=r_{max}$ completely. But due to its non-vanishing
transverse velocity orbits the black hole. For spherically symmetric black
holes, the critical value of the impact parameter corresponds to photons in
the unstable circular orbits, filling the photon sphere. The geometrical
cross-section of the photon sphere (so-called capture cross-section of the
black hole) is directly related to the critical impact parameter. For
spherically symmetric black holes, the critical impact parameter corresponds
to the area of shadow. The optical shadow around the event horizon describes
the visual boundary that light cannot escape from the event horizon by
viewers. Since the spacetime under consideration is three-dimensional, one
cannot speak about the area of shadow, instead, $b_{c} $ represents the
radius of the capture cross-section \cite{Toshmatov1}. Three-dimensional
black holes have been widely studied in the context of photon orbit and
energy emission rate \cite{Jafarzade2023,Mandal450}, bending of light \cite%
{Kala36,Gonzalez101,Panotopoulos443}, and geodesic structure \cite%
{Olivares73,Panotopoulos52,Fernando35,Cruz11}.

Taking into account the metric functions (\ref{f(r)ENMax}) and the effective
potential (\ref{Eqpotential}), $V_{\mathrm{eff}}^{\prime }(r_{ph})=0$ leads
to the following relations 
\begin{eqnarray}
m_{g}^{2}c\varepsilon r_{ph}-\frac{4q^{2}}{l^{2}}\ln \left( \frac{r_{ph}}{l}%
\right) +\frac{2q^{2}}{l^{2}}-2m_{0} &=&0,\qquad s=1,  \label{Eqrph1} \\
m_{g}^{2}c\varepsilon r_{ph}^{2}+\frac{3q}{2^{\frac{1}{4}}}-2m_{0}r_{ph}
&=&0,\qquad s=\frac{3}{4},  \label{Eqrph2} \\
\left( m_{g}^{2}c\varepsilon r_{ph}-2m_{0}\right) \left( s-1\right)
+s2^{s}\left( 1-2s\right) \left( \frac{q}{r_{ph}}\right) ^{\frac{2\left(
1-s\right) }{2s-1}} &=&0,\qquad s\in \left( \frac{1}{2},\frac{3}{4}\right)
\cup \left( \frac{3}{4},1\right) .  \label{Eqrph3}
\end{eqnarray}

Now, we examine each of the above relations separately.

\subsubsection{Three-dimensional black holes in Maxwell-massive gravity}

We consider black holes in the Maxwell-massive theory, i.e., $s=1$. Solving
Eq. (\ref{Eqrph1}) results into the following solution 
\begin{equation}
r_{ph}=l\exp \left( \frac{1}{2}-\frac{m_{0}l^{2}}{2q^{2}}-LambertW\left[ -%
\frac{m_{g}^{2}c\varepsilon l^{3}}{4q^{2}}\exp \left( \frac{q^{2}-m_{0}l^{2}%
}{2q^{2}}\right) \right] \right) .  \label{Eqrph11}
\end{equation}

\begin{figure}[!htb]
\centering
\subfloat[$ \varepsilon =1 $, $ m_{g}=0.5 $ and $ \Lambda =-0.5 $]{
		\includegraphics[width=0.305\textwidth]{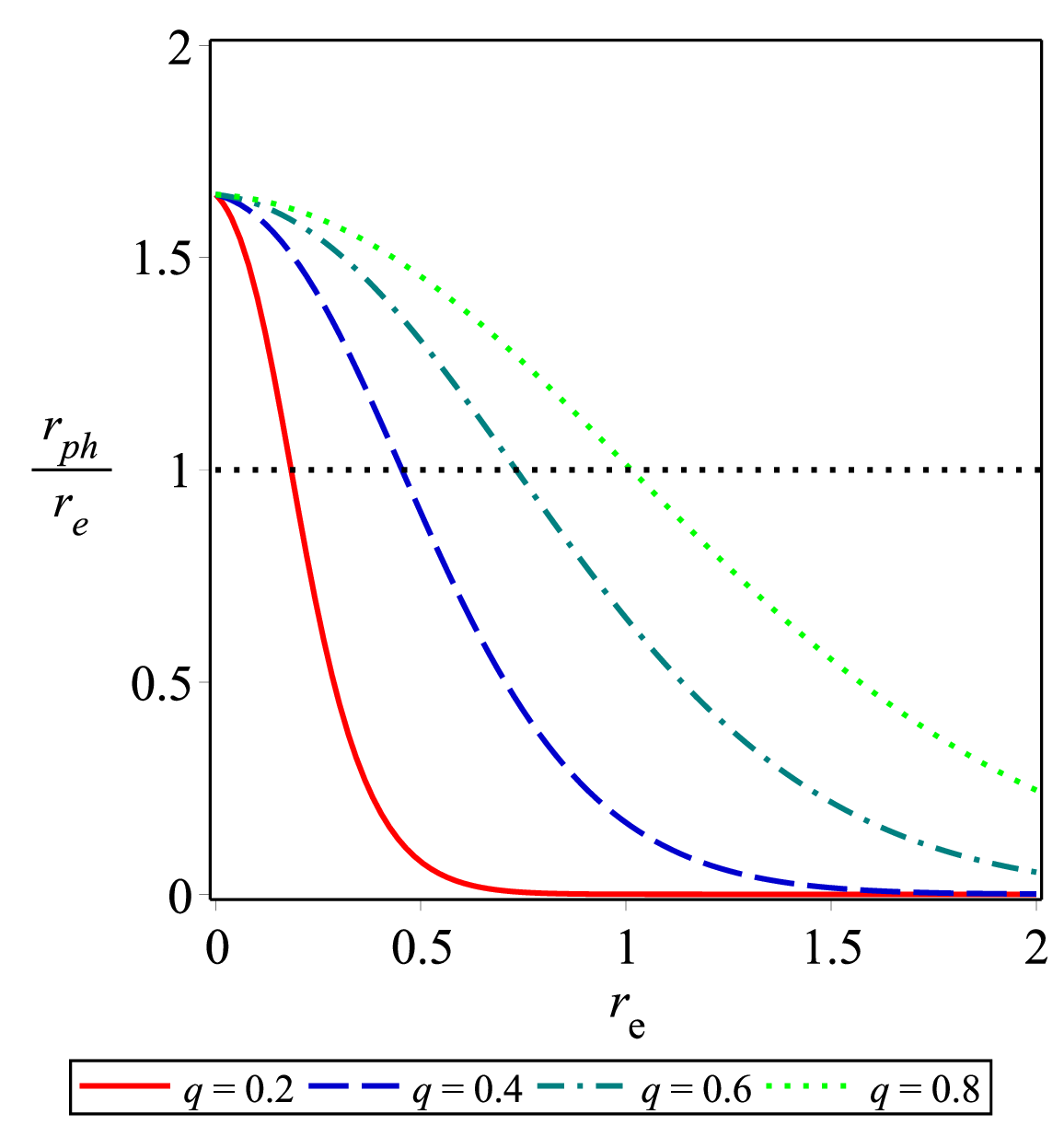}} 
\subfloat[$ q=0.2$, $\varepsilon =1 $ and $ \Lambda =-0.5 $]{
		\includegraphics[width=0.3\textwidth]{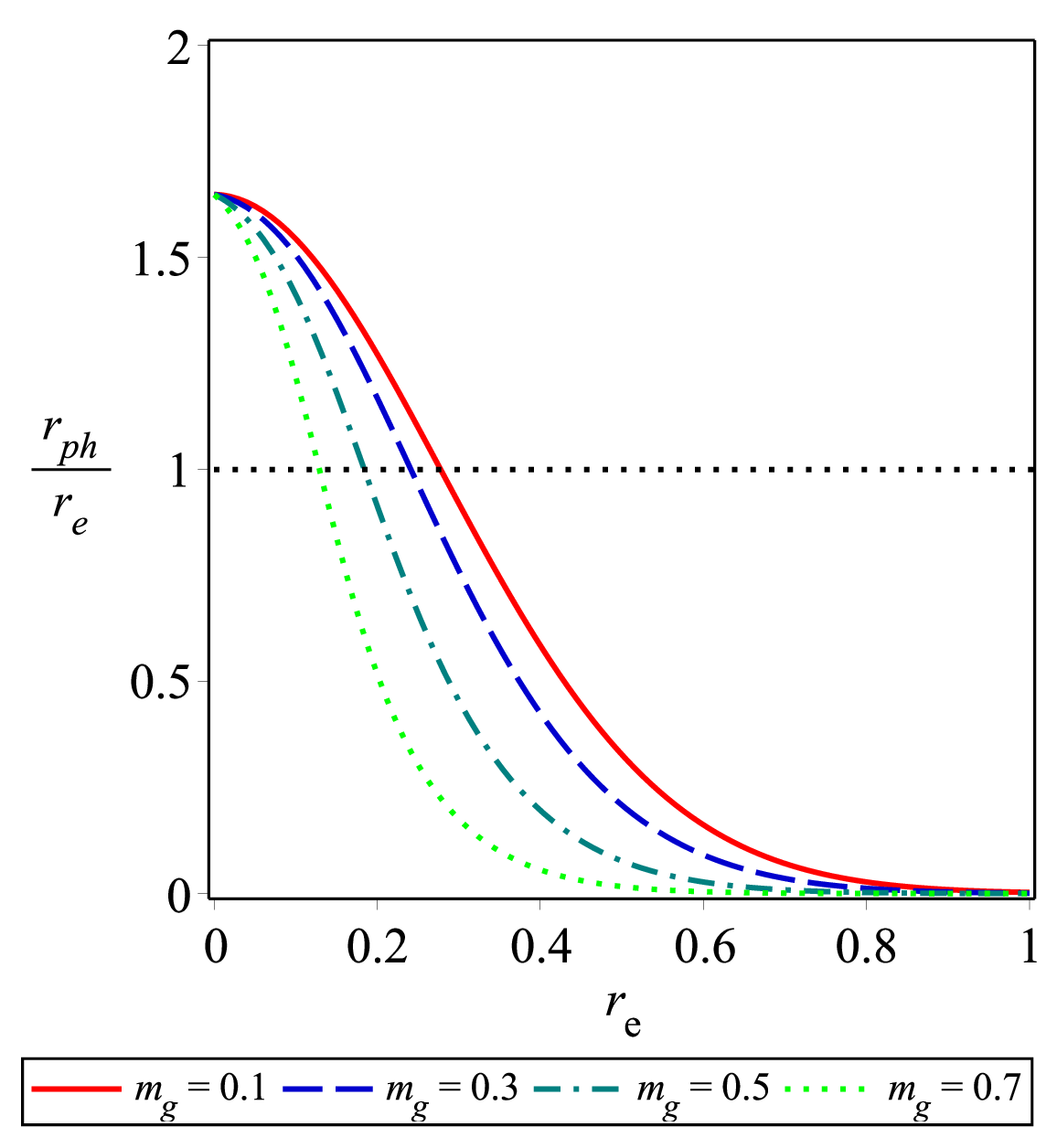}}\newline
\subfloat[$ q=0.2 $, $ m_{g}=0.5$ and $ \Lambda =-0.5 $]{
		\includegraphics[width=0.3\textwidth]{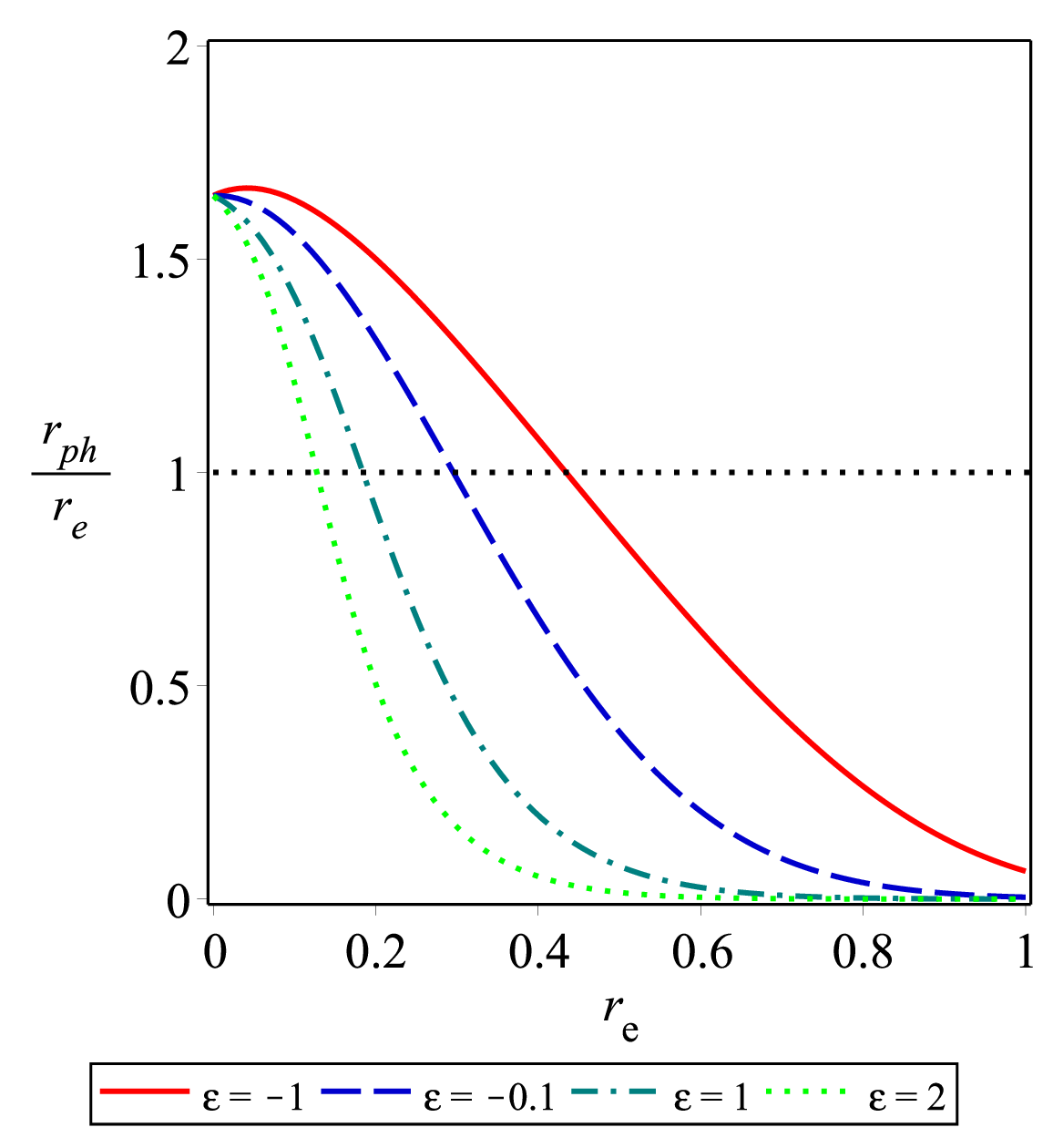}} 
\subfloat[$ q=0.2$, $\varepsilon =1 $ and $ m_{g}=0.5$]{
		\includegraphics[width=0.3\textwidth]{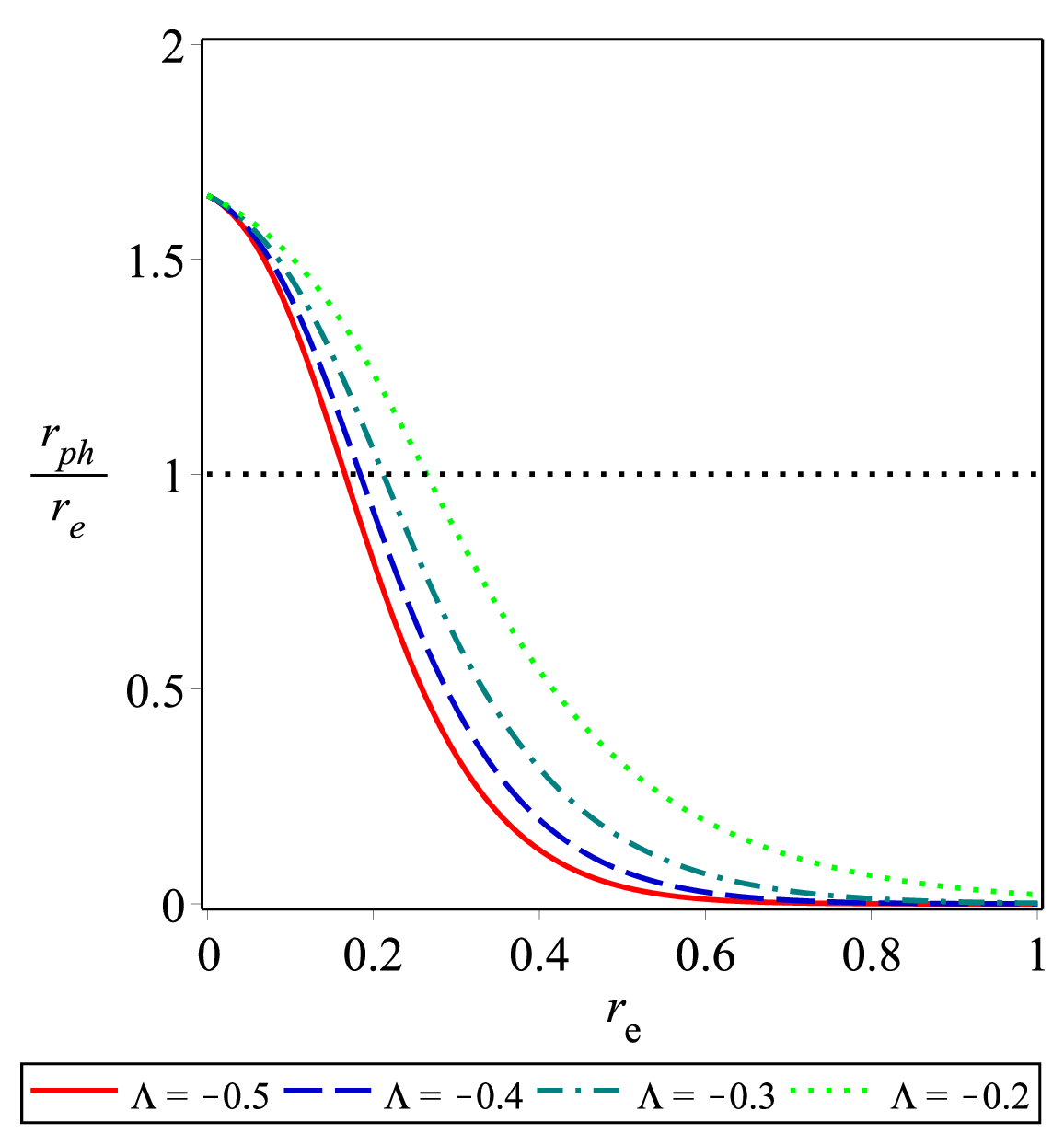}}\newline
\caption{The ratio between the photon orbit radius and the horizon radius $%
\left( \frac{r_{ph}}{r_{e}}\right) $ versus $r_{e} $ for $s=1 $, $c=l=1 $
and different values of black hole parameters.}
\label{Fig1a}
\end{figure}

As was already mentioned, by evaluating the metric function on the
horizon ($g\left(r=r_{e}\right) =0$), one can obtain the parameter $m_{0}$
as a function of $r_{e}$. Inserting the obtained $m_{0}$ into Eq. (\ref%
{Eqrph11}), $r_{ph}$ can be rewritten in terms of $r_{e}$. According to our
analysis, $r_{ph}$ has a maximum in 
\begin{equation}
r_{e,max}=\frac{m_{g}^{2}c\varepsilon l-\sqrt{m_{g}^{4}c^{2}\varepsilon
^{2}l^{2}-16\Lambda q^{2}}}{4\Lambda l},
\end{equation}
for $r_{e}<r_{e,max} $, the radius of the photon orbit is larger than the horizon radius, whereas for $r_{e}>r_{e,max} $, the photon orbit radius is smaller than the horizon radius, which is not physically acceptable. To have
a better understanding of the acceptable regions of parameters, we have
plotted Fig. \ref{Fig1a} which shows the behavior of $\frac{r_{ph}}{r_{e}}$
versus $r_{e}$. Since an acceptable optical result can be observed for $%
\frac{r_{ph}}{r_{e}}>1$ \cite{LuLyu}, only for limited regions of $r_{e}$,
this condition is satisfied. 

As a next step in the analysis, we investigate the critical impact parameter for the corresponding black hole which is obtained as \cite{Perlick1a}
\begin{equation}
b_{c}=\frac{L_{p}}{E}=\frac{r_{ph}}{\sqrt{g(r_{ph})}}.
\label{Eqrsh}
\end{equation}%
where $L_{p}$ is the critical angular momentum of the photon orbit.

According to Eq. (\ref{Eqrsh}), a real positive value of $ b_{c} $
is obtained for $g(r_{ph})>0 $. By rewriting $g(r_{ph}) $ in terms of $r_{e} 
$, one finds

\begin{equation}
\mathcal{G}=m_{g}^{2}c\varepsilon le^{\Gamma _{1}}-\Lambda l^{2}e^{2\Gamma
_{1}}+\frac{q^{2}(2\Gamma _{2}-1)}{l^{2}},  \label{Eqgnew}
\end{equation}%
where 
\begin{eqnarray}
\Gamma _{1} &=&\frac{q^{2}+2q^{2}\ln (\frac{r_{e}}{l})+\Lambda
l^{2}r_{e}^{2}-m_{g}^{2}c\varepsilon l^{2}r_{e}-2q^{2}\Gamma _{2}}{2q^{2}},
\\
&&  \notag \\
\Gamma _{2} &=&LambertW\left( -\frac{m_{g}^{2}c\varepsilon l^{3}\exp \left( 
\frac{q^{2}+2q^{2}\ln (\frac{r_{e}}{l})+\Lambda
l^{2}r_{e}^{2}-m_{g}^{2}c\varepsilon l^{2}r_{e}}{2q^{2}}\right) }{4q^{2}}%
\right) .
\end{eqnarray}

Our analysis shows that $g(r_{ph})$ is negative for all values of black hole
parameters. So, according to Eq. (\ref{Eqrsh}), the critical impact parameter is imaginary, indicating that an acceptable optical behavior cannot be observed
for three-dimensional black holes in the Maxwell-massive theory of gravity.

\subsubsection{Three-dimensional black holes in conformal invariant
Maxwell-massive gravity}

For the second case, we examine Eq. (\ref{Eqrph2}) to study the radius of
the photon orbit for three-dimensional charged AdS black holes in massive
gravity for conformally invariant Maxwell ($s=\frac{3}{4}$). Solving the
equation (\ref{Eqrph2}), one can obtain the photon orbit radius in the
following form 
\begin{equation}
r_{ph}=\frac{m_{0}+\sqrt{m_{0}^{2}-\frac{3m_{g}^{2}c\varepsilon q}{2^{1/4}}}%
}{m_{g}^{2}c\varepsilon }.  \label{Eqrph4}
\end{equation}

To investigate the ratio $\frac{r_{ph}}{r_{e}}$, we need to determine the
horizon radius, which is the root of the metric function. Our analysis shows
that the metric function can admit up to three roots as

\begin{eqnarray}
r^{(1)} &=&\frac{2\sqrt{-\Psi _{5}}}{\sqrt{3}}\sin \left[ \frac{1}{3}\sin
^{-1}\left( \frac{3\sqrt{3}\Psi _{6}}{2(\sqrt{-\Psi _{5}})^{3}}\right) %
\right] +\frac{m_{g}^{2}c\varepsilon }{3\Lambda }, \\
&&  \notag \\
r^{(2)} &=&-\frac{2\sqrt{-\Psi _{5}}}{\sqrt{3}}\sin \left[ \frac{1}{3}\sin
^{-1}\left( \frac{3\sqrt{3}\Psi _{6}}{2(\sqrt{-\Psi _{5}})^{3}}\right) +%
\frac{\pi }{3}\right] +\frac{m_{g}^{2}c\varepsilon }{3\Lambda }, \\
&&  \notag \\
r^{(3)} &=&\frac{2\sqrt{-\Psi _{5}}}{\sqrt{3}}\cos \left[ \frac{1}{3}\sin
^{-1}\left( \frac{3\sqrt{3}\Psi _{6}}{2(\sqrt{-\Psi _{5}})^{3}}\right) +%
\frac{\pi }{6}\right] +\frac{m_{g}^{2}c\varepsilon }{3\Lambda },
\label{Eqroot3}
\end{eqnarray}%
in which 
\begin{eqnarray}
\Psi _{5} &=&-\frac{m_{g}^{4}c^{2}\varepsilon ^{2}}{3\Lambda ^{2}}+\frac{%
m_{0}}{\Lambda }, \\
&&  \notag \\
\Psi _{6} &=&-\frac{2m_{g}^{6}c^{3}\varepsilon ^{3}}{27\Lambda ^{3}}+\frac{%
m_{g}^{2}c\varepsilon m_{0}}{3\Lambda ^{2}}-\frac{q}{2^{\frac{1}{4}}\Lambda }%
.  \label{spq}
\end{eqnarray}

According to our analysis, $r^{(2)}$ is always negative, and $r^{(3)}$ given
by Eq. (\ref{Eqroot3}) is the largest positive root. 
\begin{table*}[htb!]
\caption{The event horizon ($r_{e}$), photon orbit radius ($r_{ph}$) and
critical impact parameter ($b_{c}$) for the variation of $m_{g}$, $q$, $\protect%
\varepsilon $ and $\Lambda$ for $c=m_{0} =1$ and $s=\frac{3}{4} $.}
\label{tableI}\centering 
\begin{tabular}{||c|c|c|c|c||}
\hline
{\footnotesize $m_{g}$ \hspace{0.3cm}} & \hspace{0.3cm}$0.53$ \hspace{0.3cm}
& \hspace{0.3cm} $0.6$\hspace{0.3cm} & \hspace{0.3cm} $0.7$\hspace{0.3cm} & 
\hspace{0.3cm}$0.9$\hspace{0.3cm} \\ \hline
$r_{e}$ ($\Lambda =-0.01 $, $q=0.2 $, $\varepsilon=2$) & $1.5436$ & $1.1702$
& $0.7991$ & $0.3+0.09I$ \\ \hline
$r_{ph}$ ($\Lambda =-0.01 $, $q=0.2 $, $\varepsilon=2$) & $3.2867 $ & $%
2.4971 $ & $1.7459$ & $0.8810$ \\ \hline
$b_{c}$ ($\Lambda =-0.01 $, $q=0.2 $, $\varepsilon=2$) & $3.2774 $ & $%
2.5926 $ & $1.9074$ & $1.1136$ \\ \hline
$r_{ph}>r_{e}$ & \checkmark & \checkmark & \checkmark & $\times$ \\ \hline
$b_{c}>r_{ph}$ & $\times$ & \checkmark & \checkmark & \checkmark \\ 
\hline\hline
&  &  &  &  \\ 
{\footnotesize $q$ \hspace{0.3cm}} & \hspace{0.3cm} $0.04$ \hspace{0.3cm} & 
\hspace{0.3cm} $0.2$\hspace{0.3cm} & \hspace{0.3cm} $0.3$\hspace{0.3cm} & 
\hspace{0.3cm} $0.4$\hspace{0.3cm} \\ \hline
$r_{e}$ ($\Lambda =-0.01 $, $m_{g}=0.7 $, $\varepsilon=2$) & $0.9755 $ & $%
0.7991$ & $0.5453$ & $0.5+0.28I$ \\ \hline
$r_{ph}$ ($\Lambda =-0.01 $, $m_{g}=0.7 $, $\varepsilon=2$) & $1.9890 $ & $%
1.7459$ & $1.5390$ & $1.1279$ \\ \hline
$b_{c}$ ($\Lambda =-0.01 $, $m_{g}=0.7 $, $\varepsilon=2$) & $1.9833 $ & $%
1.9074$ & $1.8449$ & $1.7481$ \\ \hline
$r_{ph}>r_{e}$ & \checkmark & \checkmark & \checkmark & $\times$ \\ \hline
$b_{c}>r_{ph}$ & $\times$ & \checkmark & \checkmark & \checkmark \\ 
\hline\hline
&  &  &  &  \\ 
{\footnotesize $\varepsilon$ \hspace{0.3cm}} & \hspace{0.3cm}$-0.5$ \hspace{
0.3cm} & \hspace{0.3cm} $1.1$\hspace{0.3cm} & \hspace{0.3cm} $1.5$\hspace{
0.3cm} & \hspace{0.3cm}$3.4$\hspace{0.3cm} \\ \hline
$r_{e}$ ($\Lambda =-0.01 $, $m_{g}=0.7 $, $q=0.2 $) & $28.0443$ & $1.6136$ & 
$1.1425$ & $0.3+0.07I$ \\ \hline
$r_{ph}$ ($\Lambda =-0.01 $, $m_{g}=0.7 $, $q=0.2 $) & $-8.4081 $ & $3.4383$
& $2.4397$ & $1.2228$ \\ \hline
$b_{c}$ ($\Lambda =-0.01 $, $m_{g}=0.7 $, $q=0.2 $) & $-6.3614$ & $3.4037$
& $2.5413$ & $1.2481$ \\ \hline
$r_{ph}>r_{e}$ & $\times$ & \checkmark & \checkmark & $\times$ \\ \hline
$b_{c}>r_{ph}$ & $\times$ & $\times$ & \checkmark & \checkmark \\ 
\hline\hline
&  &  &  &  \\ 
{\footnotesize $\Lambda$ \hspace{0.3cm}} & \hspace{0.3cm}$-0.01$ \hspace{
0.3cm} & \hspace{0.3cm} $-0.04$\hspace{0.3cm} & \hspace{0.3cm} $-0.06$ 
\hspace{0.3cm} & \hspace{0.3cm}$-0.07$\hspace{0.3cm} \\ \hline
$r_{e}$ ($\varepsilon=2 $, $m_{g}=0.7 $, $q=0.2 $) & $0.7991 $ & $0.7743$ & $%
0.7590$ & $0.7517$ \\ \hline
$r_{ph}$ ($\varepsilon=2 $, $m_{g}=0.7 $, $q=0.2 $) & $1.7459 $ & $1.7459$ & 
$1.7459$ & $1.7459$ \\ \hline
$b_{c}$ ($\varepsilon=2 $, $m_{g}=0.7 $, $q=0.2 $) & $1.9074 $ & $1.8111$ & 
$1.7545$ & $1.7218$ \\ \hline
$r_{ph}>r_{e}$ & \checkmark & \checkmark & \checkmark & \checkmark \\ \hline
$b_{c}>r_{ph}$ & \checkmark & \checkmark & \checkmark & $\times$ \\ 
\hline\hline
\end{tabular}%
\end{table*}
Inserting Eq. (\ref{Eqrph4}) into Eq. (\ref{Eqrsh}), we can calculate the
critical impact parameter. To have acceptable optical behavior, we need to
examine the condition $r_{e}<r_{ph}<b_{c}$ where $r_{e}$ is the horizon
radius. Several values of $r_{e}$, $r_{ph}$, and $b_{c}$ listed in Table. %
\ref{tableI}. It can be seen that the increase of the electric charge,
graviton mass, and parameter $\varepsilon $ lead to an imaginary event
horizon. Also, for small values of these parameters, $  b_{c}$ is
smaller than the photon orbit radius, which is physically not acceptable.
This shows that an acceptable optical result can be obtained only for
limited regions of these three parameters. A remarkable point regarding the
parameter $\varepsilon $ is that for negative values of this parameter, both
photon orbit radius and critical impact parameter are negative, which is a non-physical result. Regarding the effect of the cosmological constant on $  r_{ph}$ and $  b_{c}$, we notice that an acceptable optical behavior can be observed only for small values of $|\Lambda |$. Notably, there are some acceptable optical behaviors for suitable values of $m_{g}$, $%
q$, and $\varepsilon $.

\subsubsection{Three-dimensional black holes in PM-massive gravity}

We want to investigate the optical properties of the corresponding black
hole for the total value of $s$. Nevertheless, according to the equation, we
cannot find an exact solution for the total $s$. For this purpose, we must
consider a value for $s$. Here we are interested in considering $s=\frac{4}{5%
}$. Considering $s=\frac{4}{5}$ in Eq. (\ref{Eqrph3}), we have

\begin{equation}
5m_{g}^{2}c\varepsilon r_{ph}^{5/3}-10m_{0}r_{ph}^{2/3}+12\times
2^{4/5}q^{2/3}=0,
\end{equation}

We can find the photon orbit radius by solving the above equation. Since
this equation is complicated to solve analytically, we employ numerical
methods to obtain the horizon radii, the photon orbit radius, and the
critical impact parameter. In this regard, we list several values of these three quantities in table \ref{tableII}. We see
that an acceptable optical result is obtained only for limited regions of
the electric charge, graviton mass, and parameter $\varepsilon $ since some
constraints are imposed on these parameters due to the imaginary event
horizon. Regarding the cosmological constant, just in a very low curvature
background, the photon orbit radius would be smaller than the critical impact parameter, which is physically acceptable. From this table, it can also be seen that all parameters have a decreasing effect on the event horizon, photon orbit
radius, and critical impact parameter.

\begin{table*}[htb!]
\caption{The event horizon ($r_{e}$), photon orbit radius ($r_{ph}$) and
critical impact parameter ($b_{c}$) for the variation of $m_{g}$, $q$, $\protect%
\varepsilon $ and $\Lambda$ for $c=m_{0} =1$ and $s=\frac{4}{5} $.}
\label{tableII}\centering 
\begin{tabular}{||c|c|c|c|c||}
\hline
{\footnotesize $m_{g}$ \hspace{0.3cm}} & \hspace{0.3cm}$0.49$ \hspace{0.3cm}
& \hspace{0.3cm} $0.5$\hspace{0.3cm} & \hspace{0.3cm} $0.7$\hspace{0.3cm} & 
\hspace{0.3cm}$0.8$\hspace{0.3cm} \\ \hline
$r_{e}$ ($\Lambda =-0.01 $, $q=0.1 $, $\varepsilon=1.5$) & $3.1041$ & $%
1.9922 $ & $0.8318$ & $0.41+0.05I$ \\ \hline
$r_{ph}$ ($\Lambda =-0.01 $, $q=0.1 $, $\varepsilon=1.5$) & $4.6568 $ & $%
4.4453$ & $1.9312$ & $1.2933$ \\ \hline
$b_{c}$ ($\Lambda =-0.01 $, $q=0.1 $, $\varepsilon=1.5$) & $4.6221 $ & $%
4.4688$ & $2.3515$ & $1.7555$ \\ \hline
$r_{ph}>r_{e}$ & \checkmark & \checkmark & \checkmark & $\times$ \\ \hline
$b_{c}>r_{ph}$ & $\times$ & \checkmark & \checkmark & \checkmark \\ 
\hline\hline
&  &  &  &  \\ 
{\footnotesize $q$ \hspace{0.3cm}} & \hspace{0.3cm} $0.09$ \hspace{0.3cm} & 
\hspace{0.3cm} $0.15$\hspace{0.3cm} & \hspace{0.3cm} $0.2$\hspace{0.3cm} & 
\hspace{0.3cm} $0.25$\hspace{0.3cm} \\ \hline
$r_{e}$ ($\Lambda =-0.01 $, $m_{g}=0.5 $, $\varepsilon=1.5$) & $2.0335 $ & $%
1.7787$ & $1.5266$ & $1.0+0.12I$ \\ \hline
$r_{ph}$ ($\Lambda =-0.01 $, $m_{g}=0.5 $, $\varepsilon=1.5$) & $4.5140$ & $%
4.1066$ & $3.7562$ & $3.3634$ \\ \hline
$b_{c}$ ($\Lambda =-0.01 $, $m_{g}=0.5 $, $\varepsilon=1.5$) & $4.4877 $ & $%
4.3748$ & $4.2765$ & $4.1672$ \\ \hline
$r_{ph}>r_{e}$ & \checkmark & \checkmark & \checkmark & $\times$ \\ \hline
$b_{c}>r_{ph}$ & $\times$ & \checkmark & \checkmark & \checkmark \\ 
\hline\hline
&  &  &  &  \\ 
{\footnotesize $\varepsilon$ \hspace{0.3cm}} & \hspace{0.3cm}$-0.5$ \hspace{
0.3cm} & \hspace{0.3cm} $1.4$\hspace{0.3cm} & \hspace{0.3cm} $2.0$\hspace{
0.3cm} & \hspace{0.3cm}$3.8$\hspace{0.3cm} \\ \hline
$r_{e}$ ($\Lambda =-0.01 $, $m_{g}=0.5 $, $q=0.1 $) & $17.8308$ & $2.1457$ & 
$1.4264$ & $0.41+0.03I$ \\ \hline
$r_{ph}$ ($\Lambda =-0.01 $, $m_{g}=0.5 $, $q=0.1 $) & $0.2938 $ & $4.8118$
& $3.1646$ & $1.3162$ \\ \hline
$b_{c}$ ($\Lambda =-0.01 $, $m_{g}=0.5 $, $q=0.1 $) & $-0.5633$ & $4.7317$
& $3.4547$ & $1.7766$ \\ \hline
$r_{ph}>r_{e}$ & $\times$ & \checkmark & \checkmark & $\times$ \\ \hline
$b_{c}>r_{ph}$ & $\times$ & $\times$ & \checkmark & \checkmark \\ 
\hline\hline
&  &  &  &  \\ 
{\footnotesize $\Lambda$ \hspace{0.3cm}} & \hspace{0.3cm}$-0.005$ \hspace{
0.3cm} & \hspace{0.3cm} $-0.01$\hspace{0.3cm} & \hspace{0.3cm} $-0.012$ 
\hspace{0.3cm} & \hspace{0.3cm}$-0.015$\hspace{0.3cm} \\ \hline
$r_{e}$ ($\varepsilon=1.5 $, $m_{g}=0.5 $, $q=0.1 $) & $2.0531 $ & $1.9922$
& $1.9695$ & $0.7311$ \\ \hline
$r_{ph}$ ($\varepsilon=1.5 $, $m_{g}=0.5 $, $q=0.1 $) & $4.4453 $ & $4.4453$
& $4.4453$ & $4.4453$ \\ \hline
$b_{c}$ ($\varepsilon=1.5 $, $m_{g}=0.5 $, $q=0.1 $) & $4.7102 $ & $4.4688$
& $4.3822$ & $4.2612$ \\ \hline
$r_{ph}>r_{e}$ & \checkmark & \checkmark & \checkmark & \checkmark \\ \hline
$b_{c}>r_{ph}$ & \checkmark & \checkmark & $\times$ & $\times$ \\ 
\hline\hline
\end{tabular}%
\end{table*}

\subsection{Energy emission rate}

\label{SecVB}

In this subsection, we are interested in studying the
associated energy emission rate. It has been known that  
the absorption cross-section oscillates around a limiting constant value $%
\sigma _{lim}$ at very high energies which is defined in the following form for an arbitrary
dimensional spacetime \cite{Wei1ab}
\begin{equation}
\sigma _{lim}=\frac{\pi ^{\frac{d-2}{2}}b_{c}^{d-2}}{\Gamma (\frac{d}{2})}.
\label{Eqsigma}
\end{equation}

The energy emission rate for three-dimensional spacetime is expressed as \cite{Belhaj1ab}
\begin{equation}
\frac{d^{2}E(\omega )}{dtd\omega }=\frac{4\pi ^{2}\omega ^{2}b_{c}}{e^{%
\frac{\omega }{T}}-1},  \label{Eqemission}
\end{equation}%
in which $\omega $ is the emission frequency, and $T$ denotes the Hawking
temperature.

\textbf{Conformal invariant Maxwell case:\ }for conformal invariant Maxwell
field, we have to consider $s=\frac{3}{4}$. In this case, Hawking's
temperature is calculated as 
\begin{equation}
T=\frac{1}{4\pi }\left( m_{g}^{2}c\varepsilon -2\Lambda r_{e}-\frac{q}{%
2^{1/4}r_{e}^{2}}\right) .  \label{EqeTH}
\end{equation}

\begin{figure}[!htb]
\centering
\subfloat[$m_{g}=0.7$, $\varepsilon =2 $ and $ \Lambda =-0.01
$]{\includegraphics[width=0.29\textwidth]{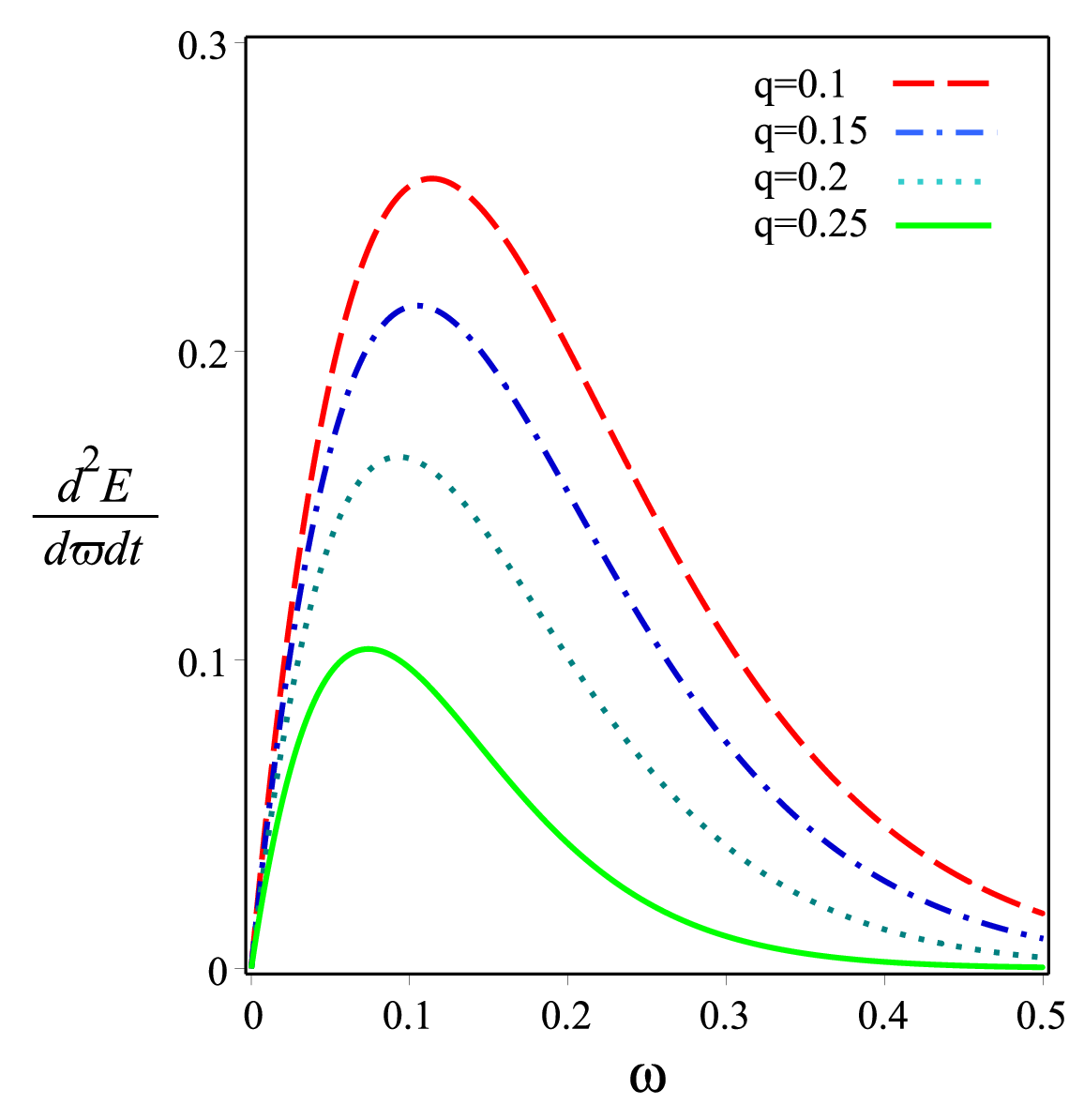}} 
\subfloat[$q=0.2$, $\varepsilon =2 $ and $ \Lambda =-0.01 $]{
		\includegraphics[width=0.3\textwidth]{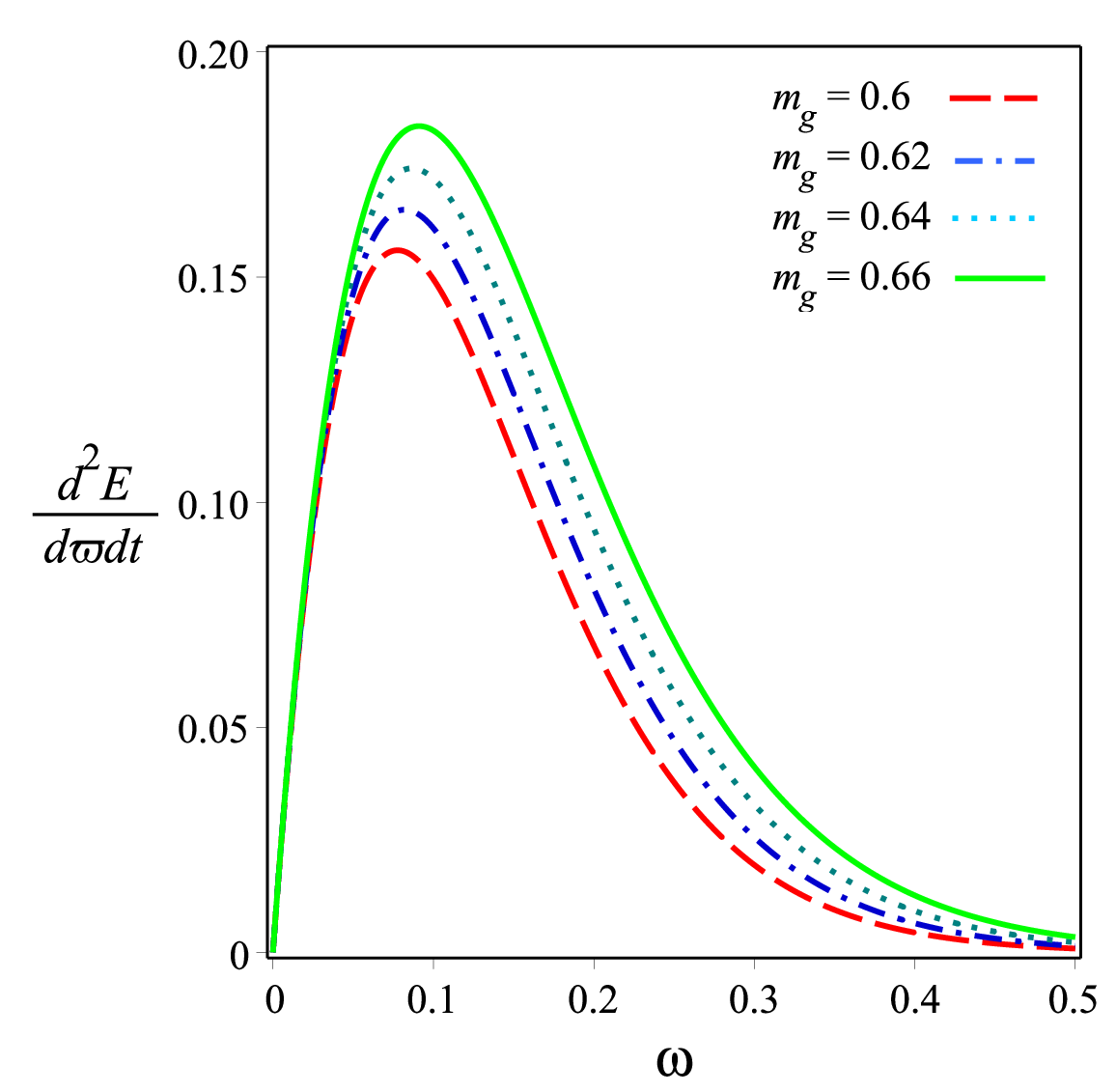}} 
\subfloat[$q=0.2$,
$\varepsilon =2 $ and $ \Lambda =-0.01 $]{
\includegraphics[width=0.3\textwidth]{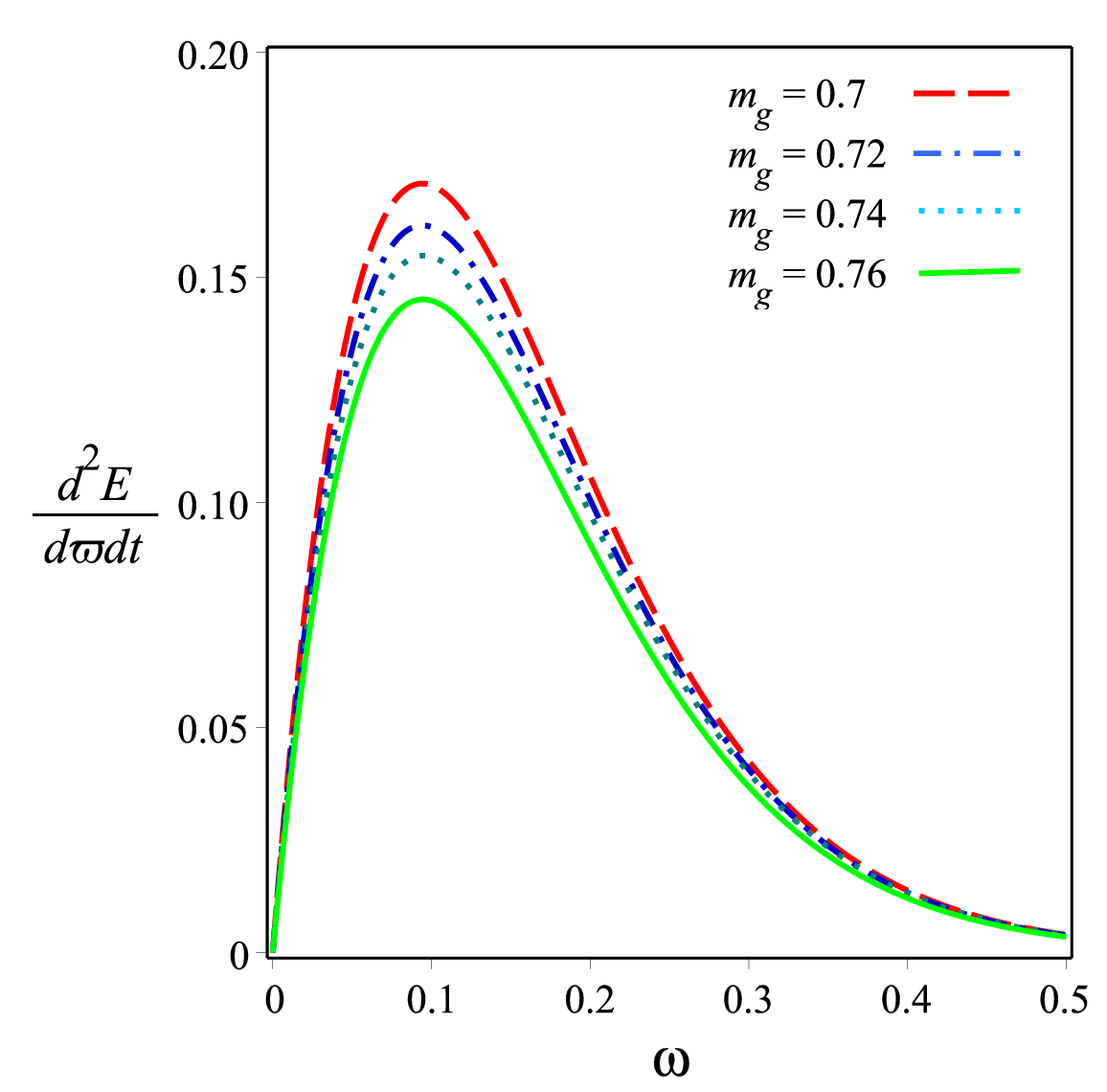}} \newline
\subfloat[$q=0.2$, $m_{g}=0.7 $ and $ \Lambda =-0.01 $]{
		\includegraphics[width=0.3\textwidth]{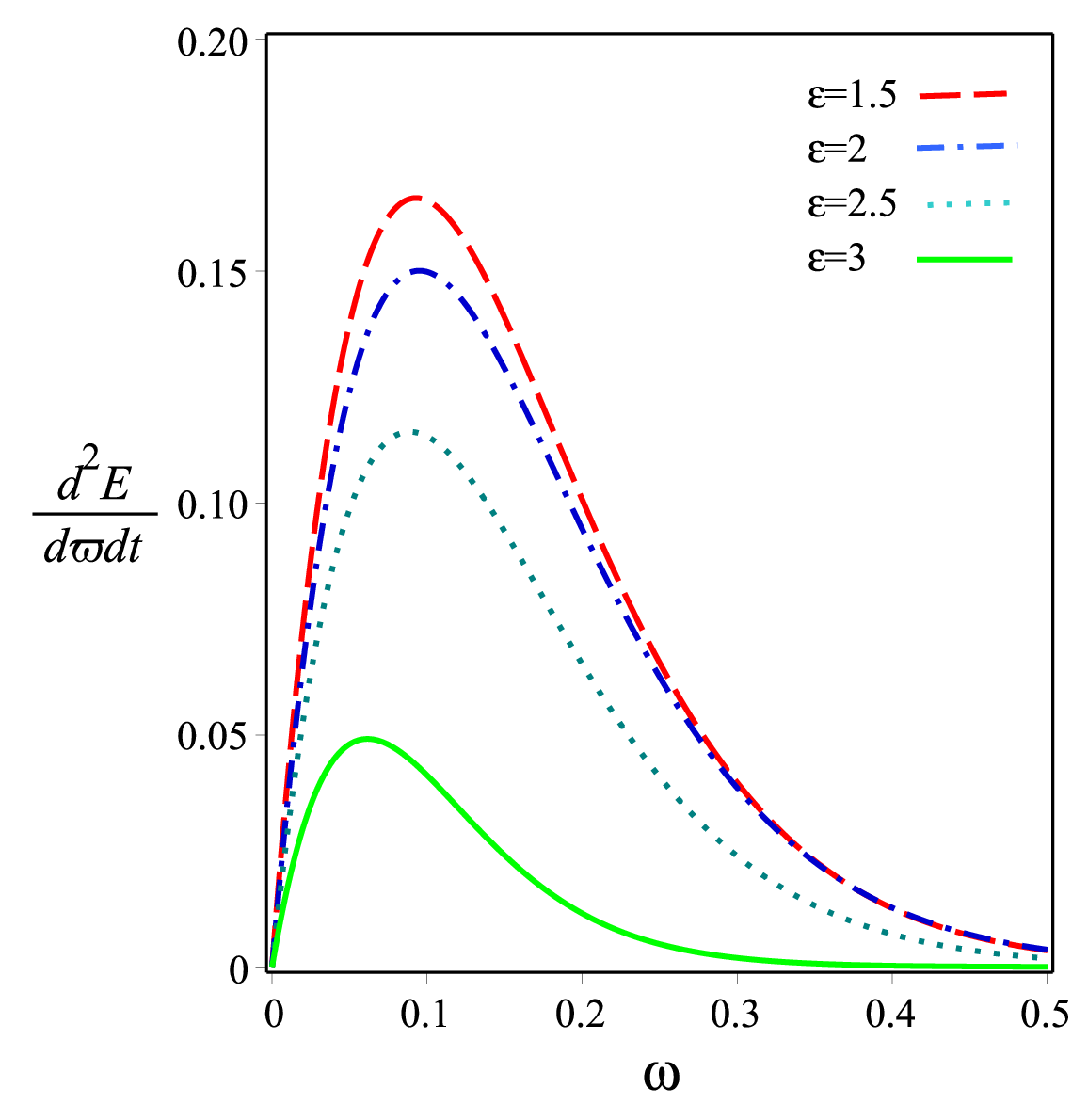}} 
\subfloat[$q=0.2$, $ m_{g}=0.7$ and $\varepsilon =2 $]{
		\includegraphics[width=0.3\textwidth]{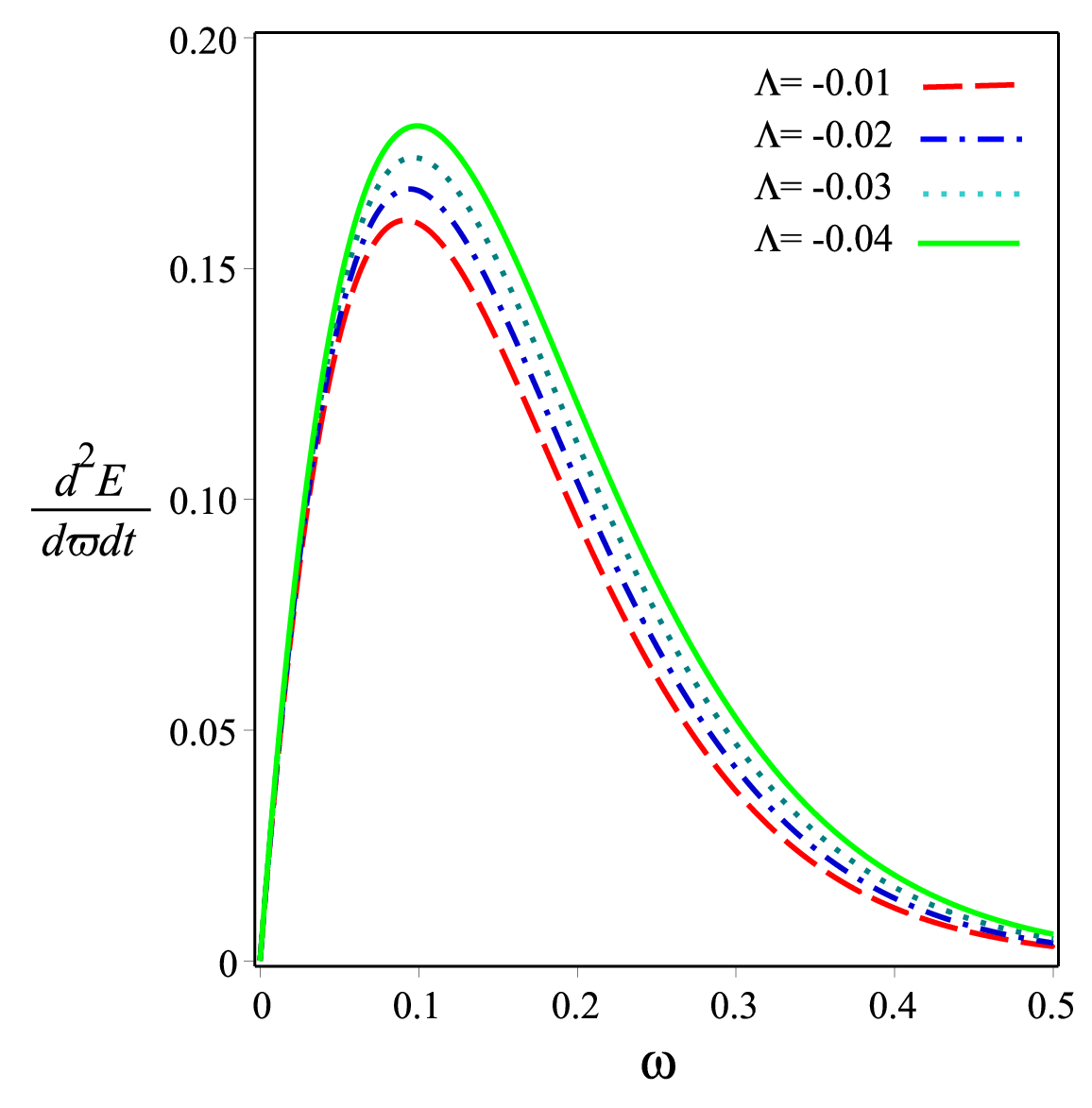}} \newline
\caption{Energy emission rate for the corresponding black hole with $s=\frac{
3}{4} $, $c=m_{0} =1$ and different values of black hole parameters.}
\label{Fig6}
\end{figure}

The qualitative behavior of the energy emission rate is illustrated in Fig. %
\ref{Fig6} as a function of $\omega $ for different values of parameters.
Looking at this figure, one can see that there exists a peak of the energy
emission rate, which decreases and shifts to the low frequency with the
increase (decrease) of the electric charge and parameter $\varepsilon $
(values of $|\Lambda |$). As one can see from Fig. \ref{Fig6}(a), the
electric charge decreases the energy emission, meaning that the evaporation
process would be slower for a black hole located in a more powerful electric
field. The effect of graviton mass on this optical quantity is a little
different such that for $0.6<m_{g}<0.7$, the energy emission rate increases
with the increase of the graviton mass, whereas for $0.7<m_{g}<0.8$ the
energy emission reduces with the growth of $m_{g}$ (see Figs. \ref{Fig6}(b)
and \ref{Fig6}(c)). Regarding the effect of $\varepsilon $, Fig. \ref{Fig6}%
(d) displays that increasing this parameter results in a decrease in energy
emission. In other words, decreasing this parameter implies a fast emission
of particles. To examine the influence of the cosmological constant, we
depict Fig. \ref{Fig6}(e), which illustrates that this parameter has an
increasing contribution to the emission rate, unlike the electric charge. In
fact, by increasing $\vert\Lambda \vert$, the energy emission rate grows.
This reveals that the evaporation process would be faster when the black
hole is located in a high curvature background. From what was expressed, one
can find that the black hole has a longer lifetime when it is located in a
low curvature background or a strong electric field.

\textbf{PM case: }for PM NED we must consider a value for $s$. Here we
consider $s=4/5$. Our investigation of the energy emission rate for $s=4/5$
shows that the effect of parameters on the energy emission is similar to the
conformal invariant Maxwell case (i.e., $s=3/4$), with this difference that
the emission rate increases by increasing the graviton mass for $%
0.5<m_{g}<0.6$ and decreases for $0.6<m_{g}<0.7$. To avoid repetition, we
omit plotting the figures.

\subsection{Deflection angle}

\label{SecVC} Here, we study the deflection angle of light using the null
geodesics method \cite{Chandrasekhar,Weinberg,Kocherlakota,WJaved}. The
total deflection $\Theta $ can be determined by the following relation 
\begin{equation}
\Theta =2\int_{b}^{\infty }\Big\vert\frac{d\varphi }{dr}\Big\vert dr-\pi ,
\label{EqDAn}
\end{equation}%
in which $b$ is the impact parameter, defined as $b\equiv L/E$. Using
equations of motion (\ref{Eqem}), we have 
\begin{equation}
\Big\vert\frac{d\varphi }{dr}\Big\vert=\frac{\dot{\varphi}}{\dot{r}}=\frac{b%
}{r^{2}}\left( 1-\frac{b^{2}g(r)}{r^{2}}\right) ^{-\frac{1}{2}}.
\label{EqDAn2b}
\end{equation}

\textbf{Conformal invariant Maxwell case: }Considering $s=3/4$ and by
inserting Eqs. (\ref{f(r)ENMax}), and (\ref{EqDAn2b}) into Eq. (\ref{EqDAn}%
), one can calculate the deflection angle as 
\begin{eqnarray}
\Theta &=&\frac{3\sqrt{2}q^{2}}{56b^{2}}+\frac{2^{\frac{3}{4}}q(1-m_{0})}{8b}%
+\frac{6-m_{0}}{3}+\frac{3\left( m_{0}^{2}+2^{\frac{3}{4}}m_{g}^{2}c%
\varepsilon q\right) }{20}-\frac{3bq\Lambda }{2^{\frac{13}{4}}}  \notag \\
&&  \notag \\
&+&\frac{bm_{g}^{2}c\varepsilon (4-m_{0})}{8}-\frac{\Lambda b^{2}(2-m_{0})}{2%
}-\frac{3\Lambda b^{3}\left( m_{g}^{2}c\varepsilon -b\Lambda \right) }{4}.
\label{EqDAn2}
\end{eqnarray}

\begin{figure}[!htb]
\centering
\subfloat[$m_{g}=0.7$, $\varepsilon =2 $ and $ \Lambda =-0.01 $]{
		\includegraphics[width=0.3\textwidth]{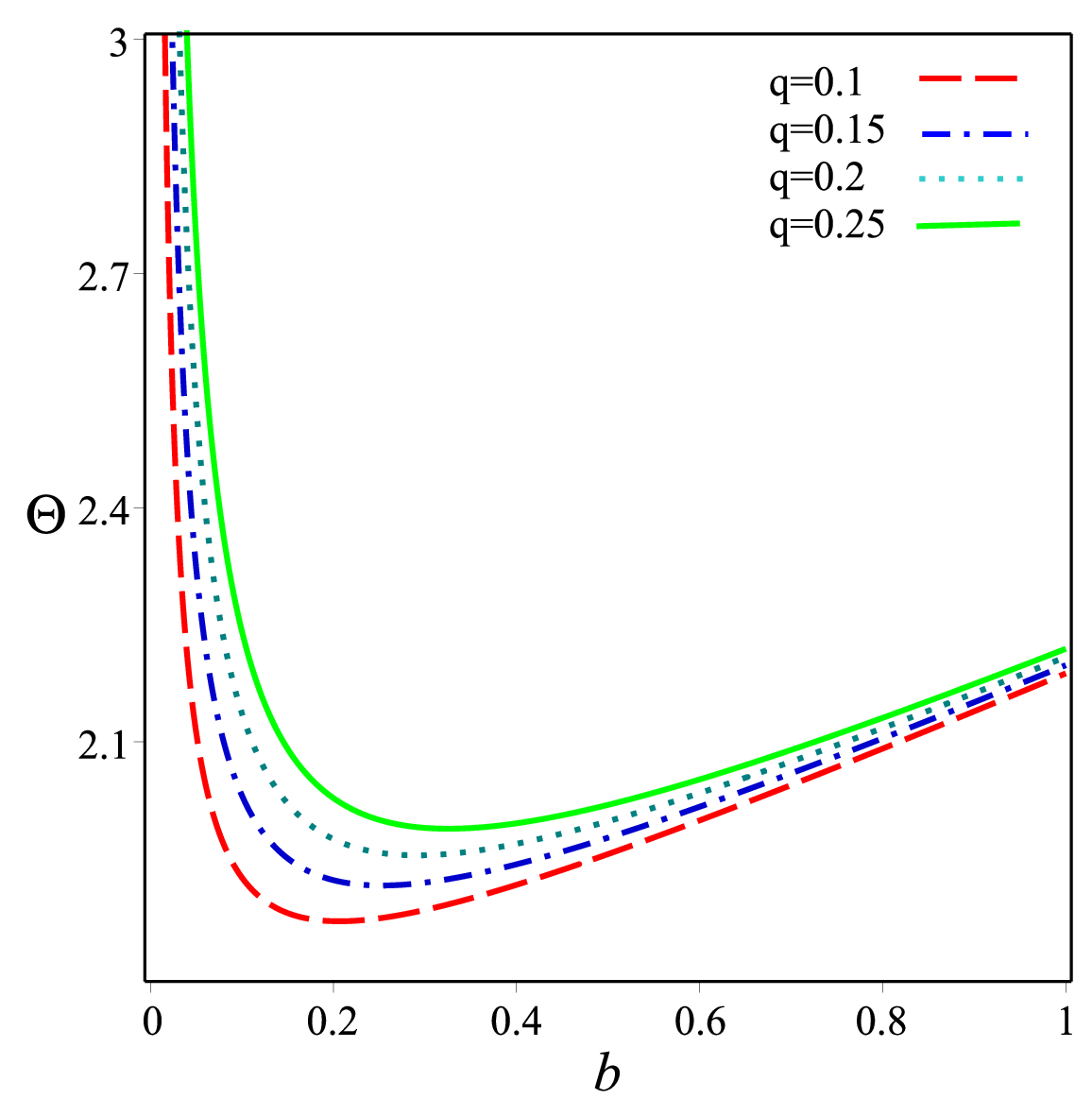}} 
\subfloat[$q=0.2$, $\varepsilon =2 $ and $ \Lambda =-0.01 $]{
		\includegraphics[width=0.3\textwidth]{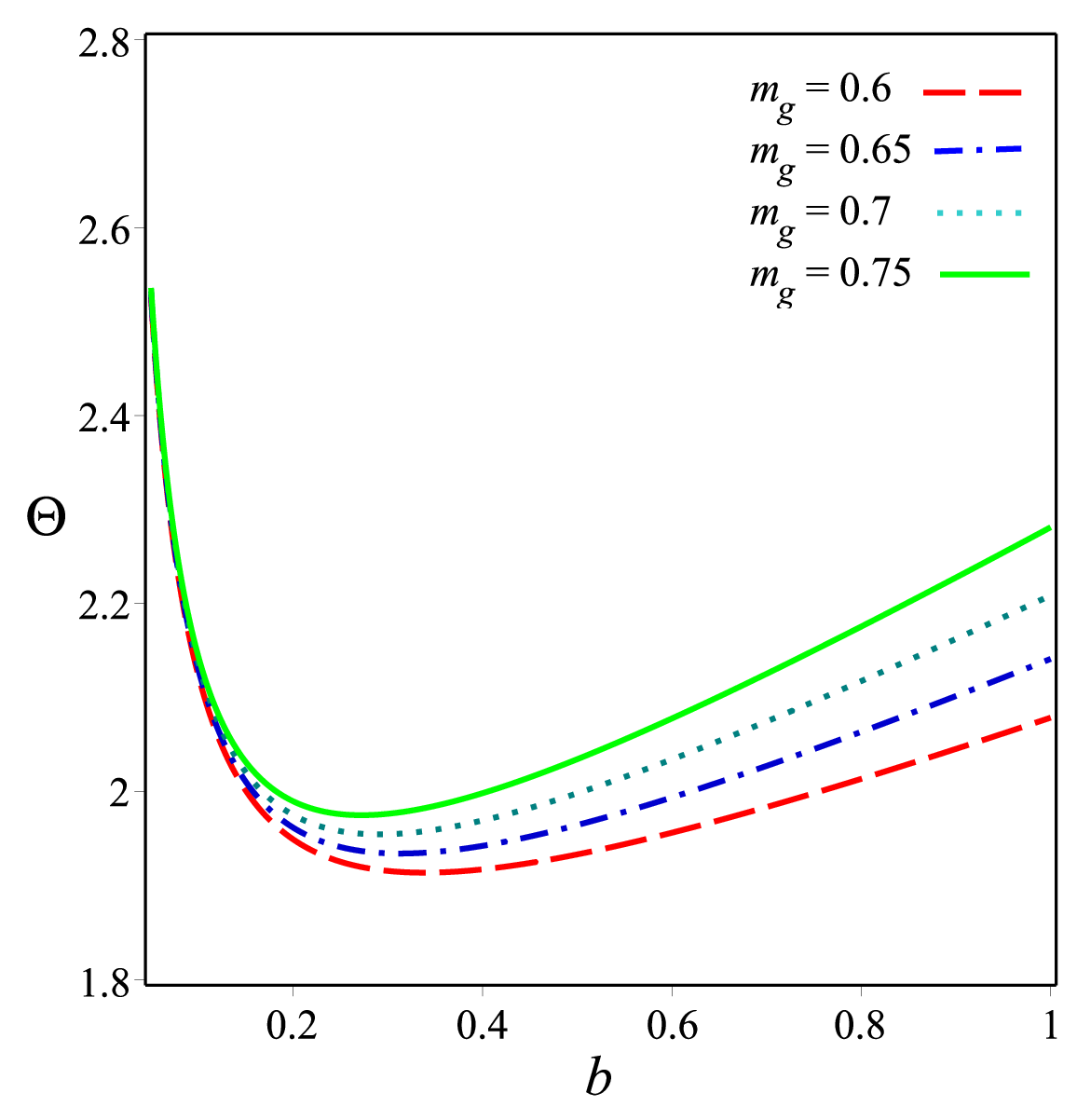}} \newline
\subfloat[$q=0.2$, $m_{g}=0.7 $ and $ \Lambda =-0.01 $]{
		\includegraphics[width=0.3\textwidth]{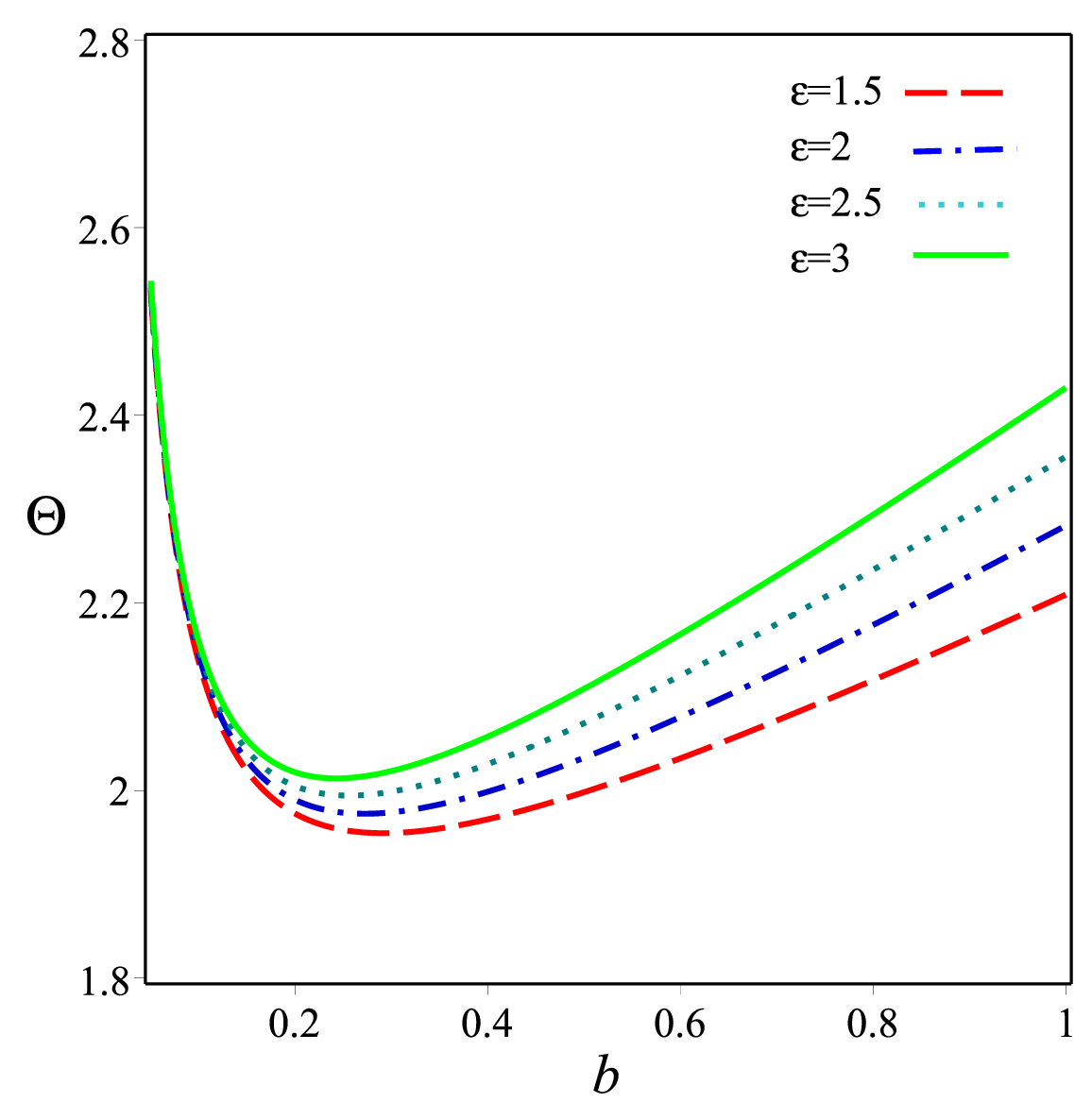}} 
\subfloat[$q=0.2$, $ m_{g}=0.7$ and $\varepsilon =2 $]{
		\includegraphics[width=0.3\textwidth]{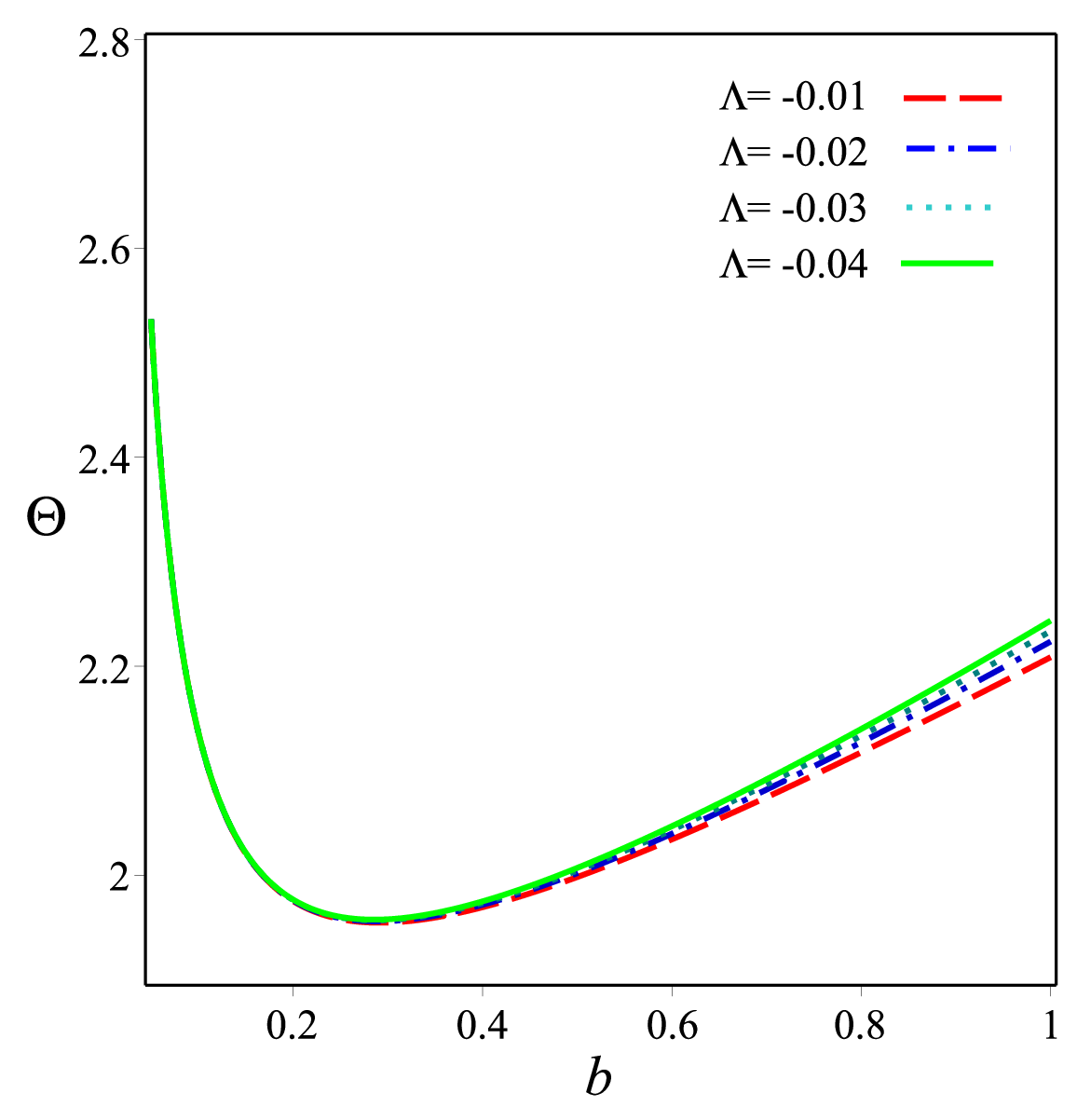}} \newline
\caption{The behavior of $\Theta $ with respect to the impact parameter $b$
for $s=\frac{3}{4} $, $c=m_{0} =1$ and different values of BH parameters.}
\label{Fig7}
\end{figure}

To show the effects of different parameters on the deflection angle, we have
depicted Fig. \ref{Fig7}, which displays the variation of the deflection
angle $\Theta $ as a function of the parameter $b$ for different values of
black hole parameters. As one can see, all curves reduce to a minimum value
with an increase of $b$ and then gradually grow as the impact parameter
increases more. In other words, they have a global minimum value, meaning
that there is a finite value of the impact parameter $b$ for which the light
deflection is very small. According to the relation $b\equiv L/E$, this
finite value is dependent on the values of angular momentum and energy of
the photon. Fig. \ref{Fig7}(a) illustrates the increasing effect of the
electric charge on the deflection angle. This shows that the light
deflection will be very high in a strong electric field. To examine the
impact of graviton mass, we have plotted Fig. \ref{Fig7}(b), indicating that
photons deflected more than their straight path in the presence of massive
gravitons. Regarding the effect of parameter $\varepsilon $, the increase of
this parameter leads to the increase of the deflection angle (see Fig. \ref%
{Fig7}(c) for more details). Studying the $\Lambda $ effect, we observe that
as $\vert\Lambda \vert$ increases, the deflection angle increases as well.
This shows that the light deflection is low in the background with lower
curvature. Comparing all the panels in Fig. \ref{Fig7}, we notice that the
effect of electric charge is notable for small values of the impact
parameter, whereas other parameters have a significant effect for large
values of $b$.

\textbf{PM case: }Since according to our analysis, for the case of $s=4/5$,
the behavior of the deflection angle under varying parameters is the same as 
$s=3/4$, we omit drawing figures.

\section{Conclusions}

\label{SecVII} In this paper, we have considered three-dimensional charged
black holes in the PM-massive gravity context. After a short introduction,
we computed the exact black hole solution and investigated several physical
properties of this black hole. Studying the thermodynamic behavior of the
solution, we examined its thermal stability and phase transition by
calculating the heat capacity in a canonical ensemble. Then, we performed an
in-depth analysis of the optical features of the corresponding black hole,
including the photon orbit radius, energy emission rate, and deflection angle, and inspected the influence of the model's parameters on the considered optical quantities.

In studying the black hole's photon orbit and critical impact parameter, we noticed that an acceptable optical behavior could not be observed for
three-dimensional black holes in the Maxwell-massive theory. Regarding the
three-dimensional charged black holes in PM-massive gravity, our analysis
showed that an admissible optical result could be obtained for special
regions of black hole parameters. Worth mentioning that such an admissible
optical result is observable only for intermediate values of the
nonlinearity parameter $s$.

Then, we studied the energy emission rate and explored the effect of black
hole parameters on the radiation process. The results indicated that as the
parameter $\varepsilon $ increases, the emission of particles around the
black hole decreases. This revealed that the radiation rate grows when the
effect of this parameter gets weaker. Studying $m_{g}$ effect showed that
depending on the value of the graviton mass, this parameter has an
increasing/decreasing contribution to the energy emission rate. Such that
for $0.6<m_{g}<0.7$, the energy emission rate increases with the increase of
the graviton mass, whereas for $0.7<m_{g}<0.8$ the energy emission reduces
with the growth of $m_{g}$. Regarding the effects of electric charge and the
cosmological constant, we noticed that the evaporation process would be slow
for a black hole located in a powerful electric field or background with
lower curvature. In other words, the lifetime of a black hole would be
longer under such conditions.

Finally, we presented a study in the context of the gravitational lensing of
light around these black holes. Depending on the values of black hole
parameters and impact parameters, photons get deflected from their straight
path and have different behaviors. For small values of the impact parameter $%
b$, the deflection angle was a decreasing function of $b$, whereas the
opposite behavior was observed for large values. It shows that there exists
a global minimum value of the impact parameter $b$, which the deflection of
light is very low for it. Relative to the impact of electric charge, we
found that it has an increasing contribution to the deflection angle $\Theta 
$. In other words, the deflection of light in a large electric charge is
very high compared to black holes located in a weak electric field. We also
noticed that the effects of graviton mass, the parameter $\varepsilon $, and
the cosmological constant on the deflection angle are similar to that of the
electric charge.

\begin{acknowledgements}

BEP thanks University of Mazandaran. Also AR acknowledges Universidad de Tarapacá for financial support.

\end{acknowledgements}

\end{document}